\documentclass[12pt,preprint]{aastex}
\usepackage{mathrsfs}
\usepackage{natbib}
\usepackage{longtable}
\usepackage{rotating}
\usepackage[usenames,dvipsnames]{color}

\newcommand{\myemail}{lqian@nao.cas.cn}
\renewcommand{\tablename}{\bf{Table}}

\def\arcsec{\hbox{$^{\prime\prime}$}}

\def\cm2{cm$^{-2}$}

\def\nh3{NH$_3$}
\def\n2h{N$_2$H$^+$}
\def\co{$^{12}$CO}
\def\13co{$^{13}$CO}
\def\c18o{C$^{18}$O}
\def\hc3n{HC$_3$N}
\def\h2{H$_2$}
\def\nh{n(H$_2$)}
\def\lp{\>\> .}
\def\lc{\>\> ,}
\def\Ms{$M_{\odot}$}

\begin{document}

\slugcomment{}

\shorttitle{$^{13}$CO Cores in Taurus} \shortauthors{Qian, Li}

\title{$^{13}$CO Cores in Taurus Molecular Cloud}

\author{Lei Qian \altaffilmark{1}, Di Li \altaffilmark{1,2,3} and  Paul F. Goldsmith \altaffilmark{4}}
\affil{} \altaffiltext{1} {National Astronomical Observatories,
Chinese Academy of Sciences, Beijing, 100012, China. Email:\myemail}
\altaffiltext{2} {Space Science Institute, Boulder, CO, Email:
ithaca.li@gmail.com}
\altaffiltext{3} {Department of Astronomy,
California Institute of Technology, CA, USA}
\altaffiltext{4} {Jet
Propulsion Laboratory, California Institute of
  Technology, Pasadena, CA, USA}

\begin{abstract}
Young stars form in molecular cores, which are dense condensations
within molecular clouds.  We have searched for molecular cores
traced by $^{13}$CO $J=1\to 0$ emission in the Taurus molecular
cloud and studied their properties. Our data set has a spatial
dynamic range (the ratio of linear map size to the pixel size) of
about 1000 and spectrally resolved velocity information, which
together allow a systematic examination of the distribution and
dynamic state of $^{13}$CO cores in a large contiguous region. We
use empirical fit to the CO and CO$_2$ ice to correct for depletion
of gas-phase CO. The $^{13}$CO core mass function ($^{13}$CO CMF)
can be fitted better with a log-normal function than with a power
law function. We also extract cores and calculate the $^{13}$CO CMF
based on the integrated intensity of $^{13}$CO and the CMF from
2MASS. We demonstrate that there exists core blending, i.e.\
combined structures that are incoherent in velocity but continuous
in column density.

The core velocity dispersion (CVD), which is the variance of the
core velocity difference $\delta v$, exhibits a power-law behavior
as a function of the apparent separation $L$:\  CVD (km/s) $\propto
L ({\rm pc})^{0.7}$. This is similar to Larson's law for the
velocity dispersion of the gas. The peak velocities of \13co\ cores
do not deviate from the centroid velocities of the ambient \co\ gas
by more than half of the line width. The low velocity dispersion
among cores, the close similarity between CVD and Larson's law, and
the small separation between core centroid velocities and the
ambient gas all suggest that molecular cores condense out of the
diffuse gas without additional energy from star formation or
significant impact from converging flows.

\end{abstract}


\section{Introduction}

Most young stars are found in dense molecular
cores~\citep{McKee2007}. There is a large volume of data concerning
dense molecular cores traced by dust emission and dust
extinction~\citep{Motte1998,Testi1998,Johnstone2000,Stanke2006,Reid2005,Reid2006,Johnstone2001,Johnstone2006,DCMF_Alves_Lombardi_Lada}.
One of the primary outcomes of these studies is the core mass
function (CMF). Because of the still unknown physical origin of the
stellar initial mass function (IMF) and its significance, emphasis
has been placed on the possible connection between the CMF and the
IMF~\citep{DCMF_Alves_Lombardi_Lada}.

The construction of an analytic form of the CMF from observational
data has largely focused on two functional forms, power law and
log-normal. The majority of past studies claim to find a power law
CMF, the shape of which resembles the Salpeter IMF~\citep{IMF} (The
upper mass limit of the original Salpeter IMF is $10 M_{\odot}$)
\begin{equation}
\frac{{\rm d}N}{{\rm d}\log M} \propto M^{-\gamma},\ \ \gamma=1.35, -0.4\le\log(M/M_{\odot})\le 1.0 \lp
\label{power_law}
\end{equation}
Such a power law CMF was found in millimeter continuum maps of the
$\rho$ Ophiuchus region by \cite{Motte1998} with the IRAM 30-meter
telescope and a series of subsequent studies in the Serpens region
\citep{Testi1998}, the $\rho$ Ophiuchus region
\citep{Johnstone2000,Stanke2006}, NGC 7538 \citep{Reid2005}, M17
\citep{Reid2006}, Orion \citep{Johnstone2001,Johnstone2006} and the
Pipe nebula \citep{DCMF_Alves_Lombardi_Lada}. \cite{Reid2006b}
studied the CMF in 11 star-forming regions and find an average power
law index of $\gamma=1.4\pm 0.1$. Recently, dust emission
observations of the Aquila rift complex with {\it Herschel}
\footnote{http://www.esa.int/SPECIALS/Herschel/index.html} reveal a
power law mass function with $\gamma=1.5\pm0.2$, for $M>2\
M_{\odot}$~\citep{Aquila_cores}, which is also consistent with the
Salpeter IMF. A similar power law index is found in some molecular
emission studies. For example, the CMF obtained from a $\rm C^{18}O$
study in the S140 region has $\gamma=1.1\pm 0.2$~\citep{S140}. On
the other hand, there are also studies that find a flatter CMF.
\cite{Kramer1998} studied seven molecular clouds L1457, MCLD
126.6+24.5, NGC 1499 SW, Orion B South, S140, M17 SW, and NGC 7538
in $^{13}$CO and C$^{18}$O, and find $\gamma$ to be between 0.6 and
0.8. \cite{core_mass_function} studied cores in the Orion molecular
cloud traced by sub-millimeter continuum and found a power law with
$\gamma=0.15\pm 0.21$. These findings of small $\gamma$ are in the
minority and do not seem to be a special result of spectroscopic
mapping. \cite{Pipe_nebula_improve} obtained the CMF from an
extinction map and used complementary C$^{18}$O observations to
examine the effects of blending of cores in dust maps. They claimed
to find a CMF with $\gamma$ similar to that of the Salpeter IMF.

At a first glance, a similarity between the CMF and Salpeter IMF
suggests a constant star formation efficiency, which is independent
of the core mass. It is crucial to note, however, these studies are
examining structures on vastly different scales of size, mass, and
density. In \cite{Reid2006}, for example, the mass of the cores
ranges from about 0.1 $M_{\odot}$ to $1.6\times 10^4\ M_{\odot}$.
Due to the large distances of many targeted regions, any "cores"
over about 500 \Ms\ are certainly unresolved, with many showing
signs of much evolved star formation, such as water masers
\citep{Wang2006} and/or compact HII regions \citep{Hofner2002}. The
observed similarity between CMF and Salpeter IMF may be explained
equally well by self-similar  cloud structures as well as a constant
star formation efficiency.

Some observations (e.g., \cite{observe_lognormal}) suggest a
log-normal form for the CMF (in the mass range of $0.1\
M_{\odot}<M<10\ M_{\odot}$),
\begin{equation}
\frac{{\rm d}N}{{\rm d}\log M} \propto \exp\left[-\frac{(\log
M-\mu)^2}{2\sigma^2}\right]\lp
 \label{lognormal}
\end{equation}
This is reminiscent of the \cite{Chabrier2003} IMF, which is also of
log-normal form. Theoretically, if the core mass depends on $n$
quantities which are random variables, the CMF would be log-normal
when $n$ is large, i.e., the core formation processes are
complicated~\citep{adams1996}. This is a result of the central limit
theorem. A log-normal distribution also arises naturally from
isothermal turbulence~\citep{Larson1973}.

It is thus of great interest to distinguish the two forms of the CMF and
obtain the key parameters associated with each form.
\cite{Form_of_CMF} show that a large sample with many cores is needed to differentiate these two forms.
Furthermore, we also emphasize here the critical need to obtain a large sample of cores in spectroscopic data.
Overlapping cores along the same line of sight can only be separated using resolved velocity information.
It is important to evaluate the
effect of such accidental alignment on the derived CMF.
A Nyquist sampled continuous spectroscopic map is also essential for
studying the  dynamical characteristics of star forming regions, such as the Core Velocity Dispersion (CVD $\equiv \langle\delta v^2\rangle^{1/2}$ see section 4.3),
of the whole core sample in one star forming region.

The Taurus molecular cloud is a nearby (with a distance of 140
pc,~\cite{Distance}) low-mass star-forming region. In this work, we
obtain a sample of cores in the $^{13}$CO data cube of this region.
We study the properties of $^{13}$CO cores in detail and compare
them with those found in the dust extinction map of the same region.
We first briefly describe the data in \S \ref{sec:data}; we present
the methods used to find $^{13}$CO cores in \S \ref{sec:methods}; we
present the observed $^{13}$CO CMF and CVD in \S \ref{sec:results};
we discuss the implications of our observations in \S
\ref{sec:discussion}. In the final section we present our
conclusions.

\section{The Data}
\label{sec:data}

In this work, we use $^{12}$CO and $^{13}$CO data in the form of a
(x,y,v) cube of the Taurus molecular cloud as observed with the 13.7
m FCRAO telescope \citep{Taurus_CO} and the 2MASS extinction map of
the same region \citep{2MASS}. The $^{12}$CO and $^{13}$CO lines
were observed simultaneously between  2003 and 2005. The map is
centered at $\alpha(2000.0)=04^h 32^m 44.6^s$,
$\delta(2000.0)=24^\circ 25' 13.08''$, with an area of $\sim 98\ \rm
deg^2$. The FWHM beam width of the telescope is $45''$ at 115 GHz.
The angular spacing (pixel size) of the resampled on the fly (OTF)
data is 20'' \citep{Goldsmith2008}, which corresponds to a physical
scale of $\approx 0.014\rm\ pc$ at a distance of $D=140\ {\rm pc}$.
There are 80 and 76 velocity channels in the $^{12}$CO and $^{13}$CO
data cube, respectively. The width of a velocity channel is $V_{\rm
ch}=0.266 \rm\ km/s$. The extinction map has a pixel size of about 5
times that of the CO data cube with, of course, no velocity
information.

\begin{figure}[htb]
\centering
\includegraphics[width=18cm]{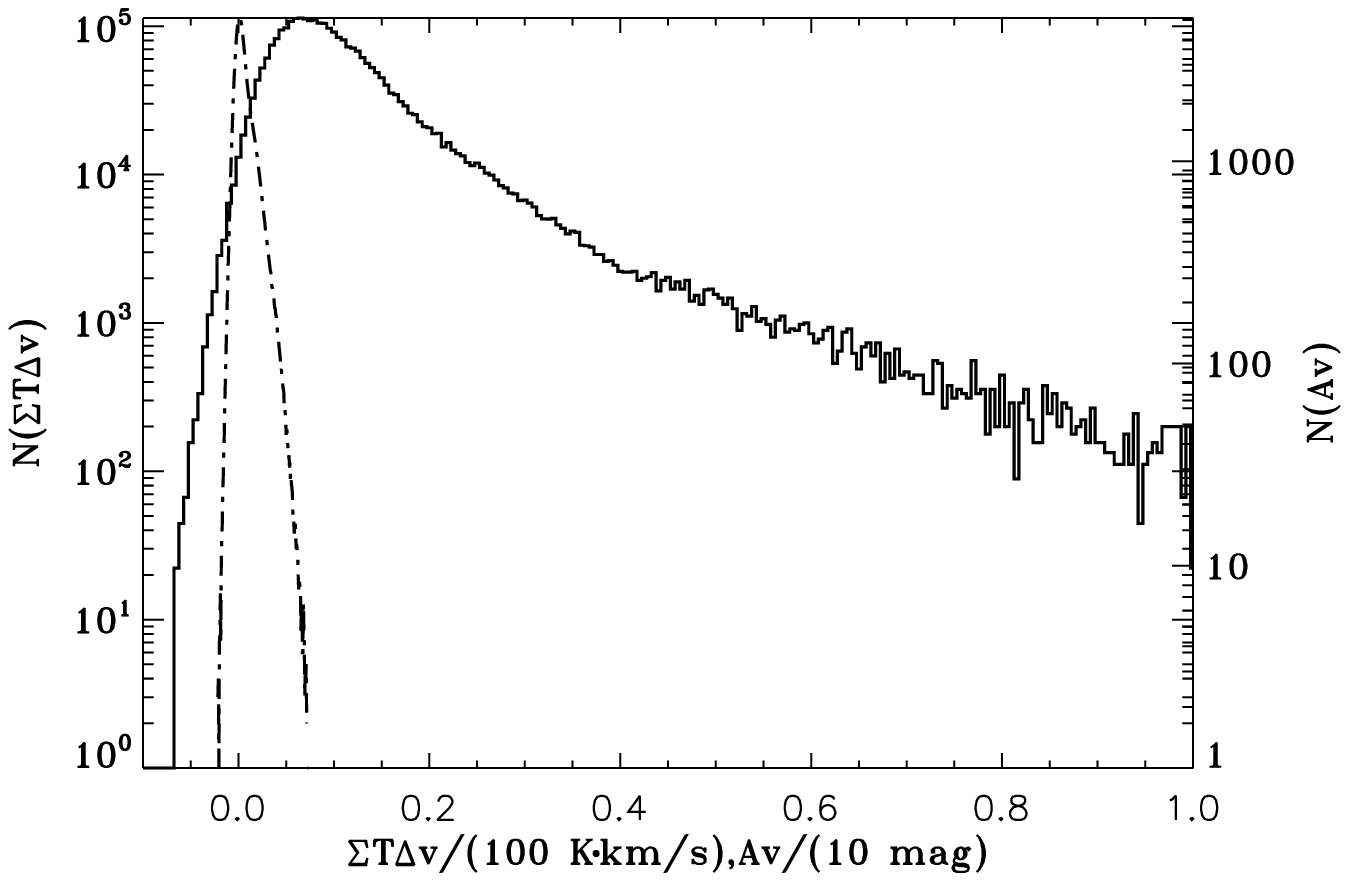}
\caption{The PDFs of $^{13}$CO integrated intensity (dot-dashed
line) and 2MASS extinction (solid line) are shown. The horizontal
axis shows the $^{13}$CO integrated intensity and the 2MASS
extinction (normalized to values that correspond to similar column
density of molecular hydrogen), while the vertical axis shows the
number of pixels with different values ({\it left:} $^{13}$CO
integrated intensity; {\it right:} 2MASS extinction).
 \label{pdf}}
\end{figure}

The probability density function (PDF) of $^{13}$CO data and that of
2MASS extinction data are shown in Figure~\ref{pdf}, from which the
difference between the two data set can be seen. The normalization
of $^{13}$CO integrated intensity and that of 2MASS extinction are
chosen so that they roughly correspond to the same column density of
molecular hydrogen.

\section{Core Extraction}
\label{sec:methods}

Cores have been empirically defined as regions with concentrated,
enhanced intensity in a data cube or a map. We empirically assume an
ellipsoidal shape for a core and use the FINDCLUMPS tool in the
CUPID package, which is a part of the starlink
software\footnote{http://starlink.jach.hawaii.edu/starlink/}. We
have tried two methods in the FINDCLUMPS tool, GAUSSCLUMPS
\citep{Stutzki1990} and CLUMPFIND \citep{Williams1994}. GAUSSCLUMPS
searches for an ellipsoid with Gaussian density profile around the
brightest peak and subsequently subtracts it from the data. It then
continues the process with the core-removed data, iterating
successively until a terminating criterion is reached. CLUMPFIND
identifies cores by drawing enclosing contours around intensity
peaks without assuming the shape of cores a priori. Unlike
GAUSSCLUMPS, CLUMPFIND cannot deconvolve overlapping cores. Our
subsequent analysis is based on GAUSSCLUMPS and we discuss the
caveats of CLUMPFIND at the end of this section.

For fitted ellipsoids of revolution, the core radius is defined as the geometrical mean of
the semi-major and the semi-minor axes
\begin{equation}
R \equiv (R_{\rm max}R_{\rm min})^{1/2}\lp \label{core_size}
\end{equation}
We take the observed size $R$ as a typical scale of a core.

Following the instructions for CUPID, we first subtract the
background by using FINDBACK. Some studies of the Taurus molecular
cloud find a characteristic length scale of about $0.5 {\rm pc}\sim
2{\rm pc}$, at which self-similarity breaks down and gravity becomes
important \citep{not_fractal}. This length scale corresponds roughly
to $35\sim 140$ pixels in the $^{13}$CO data cube, and to $7\sim 28$
pixels in the extinction map. We set the smoothing scale to 127
pixels in each axis for the $^{13}$CO data cube, and 25 pixels for
the extinction map, both are close to the upper value of the scale
at which the self-similarity breaks down.

\begin{figure}[htb]
\centering
\includegraphics[width=18cm]{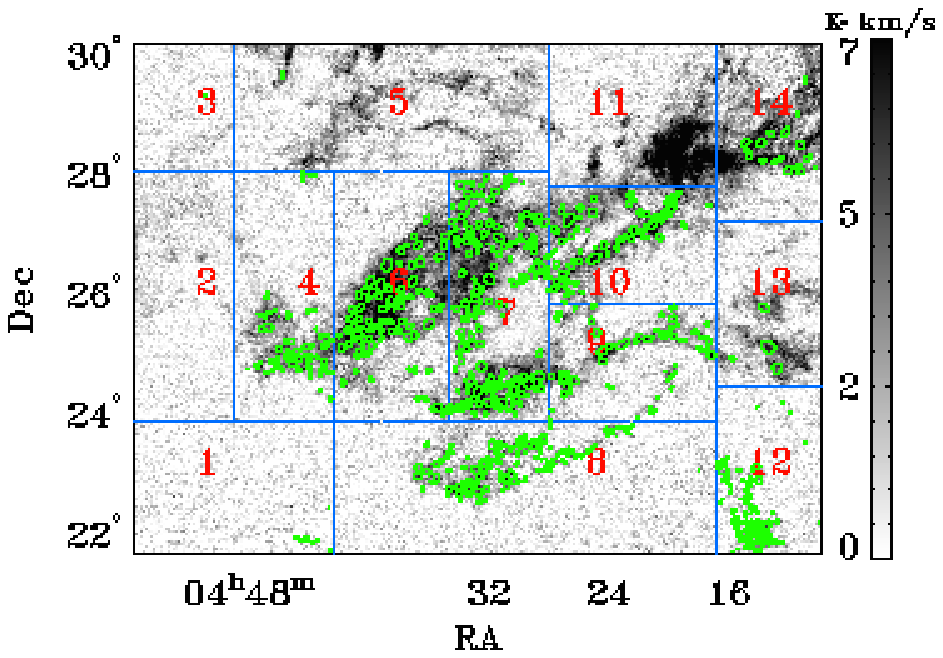}
\caption{ Image showing the total intensity of the $^{13}$CO
emission in Taurus region. The data are divided into 18 regions for
$^{13}$CO core fitting. The cores found by fitting gaussians to the
\13co\ data cube $(x,y,v)$ using GAUSSCLUMPS are shown as green
ellipsoids. Despite the strong \13co\ emission, no $^{13}$CO cores
are found in region 11, and only one $^{13}$CO core is found in
region 5. This can be understood by looking at the channel maps (see
discussion in section 3).
 \label{fig1}}
\end{figure}

After background subtraction, we use the GAUSSCLUMPS method of the
FINDCLUMPS tool to fit Gaussian components in the $^{13}$CO data
cube , which are identified as $^{13}$CO cores if their peak
intensity is larger than a threshold (see e.g. \cite{Curtis2010}),
which is set to five times of the rms noise (parameter RMS). Since
the data cube is large, running the 3D-Gaussian fitting on the whole
data cube is time-consuming. Furthermore, the background differs
substantially between different parts of the Taurus cloud. The noise
level also differs among different regions. The data cube is divided
into several sections in the x-y plane as shown in figure \ref{fig1}
to make the sizes of data sets suitable to handle, and to minimize
the variation of the background and the noise level. The fitting
parameters are given in table~\ref{gaussclumpspara}. In practice,
two adjacent regions are arranged to have an overlapping stripe of
100 pixels in width to avoid the cores touching boundaries being
missed. In a final step, the resultant catalog is checked and the
duplicate cores are removed.

FINDCLUMPS outputs the total intensity, $T_{\rm tot}$, through the
summation of all pixels in each fitted core. We then calculate the
mass of a core based on $T_{\rm tot}$. The central frequency of the
$^{13}$CO $J=1\to 0$ line $\nu$ is 110.2 GHz. The column density of
\13co\ in the upper-level ($J=1$) can be expressed
as~\citep{RadioTool}
\begin{equation}
N_{u,^{13}\rm CO}=\frac{8\pi k\nu^2}{hc^3 A_{ul}}\int T_b {\rm d}v
\lc
\end{equation}
where $k$ is Boltzmann's constant, $h$ is Planck's constant, $c$ is the speed of light, $A_{ul}$
is the spontaneous  decay rate from the upper level to the lower level, and $T_b$ is the brightness temperature.
A convenient form of this equation is
\begin{equation}
\left(\frac{N_{u,^{13}\rm CO}}{\rm cm^{-2}}\right)=3.04\times
10^{14}\int \left(\frac{T_b}{\rm K}\right) {\rm d}\left(\frac{v}{\rm
km/s}\right) \lp
\end{equation}
The total \13co\ column
density $N_{\rm tot}$ is related to the upper level column density
$N_{u}$ through \citep{Di_Li_Thesis}
\begin{equation}
N_{\rm tot, ^{13}CO}=f_{u} f_{\tau} f_b N_{u,\rm ^{13}CO} \lp
\end{equation}
In the equation above, the level correction factor $f_u$ can be
calculated analytically under the assumption of local thermal
equilibrium (LTE) as
\begin{equation}
f_{u}=\frac{Q(T_{\rm ex})}{g_u \exp\left(-\frac{h\nu}{kT_{\rm ex}}\right)} \lc
\end{equation}
where $g_u$ is the statistical weight of the upper-level. $T_{\rm
ex}$ is the excitation temperature and $Q(T_{\rm ex})=kT_{\rm ex}/B_e$ is the LTE
partition function, where $B_e$ is the rotational constant
\citep{Astrospec}. A convenient form of
the partition function is then $Q(T_{\rm ex})\approx T_{\rm ex}/2.76\rm K$.
The correction factor for opacity is defined as
\begin{equation}
f_{\tau}=\frac{\int\tau_{13}dv}{\int(1-e^{-\tau_{13}}){\rm d}v} \lc
\end{equation}
and the correction for the background
\begin{equation}
f_b=\left[1-\frac{e^{\frac{h\nu}{kT_{\rm
ex}}}-1}{e^{\frac{h\nu}{kT_{\rm bg}}}-1}\right]^{-1} \lc
\end{equation}
where $\tau_{13}$ is the opacity of the $^{13}$CO transition and
$T_{\rm bg}$ is the background temperature, assumed to be 2.7K.

The \13co\ opacity is estimated as follows. Assuming equal excitation temperatures for the two
isotopologues, the ratio of the brightness temperature of $^{12}$CO to that of $^{13}$CO
can be written as
\begin{equation}
\frac{T_{b, 12}}{T_{b, 13}}=\frac{1-e^{-\tau_{12}}}{1-e^{-\tau_{13}}} \lp
\end{equation}
Assuming $\tau_{12}\gg 1$, the opacity of $^{13}$CO can be written as
\begin{equation}
\tau_{13}=-\ln\left(1-\frac{T_{b,13}}{T_{b,12}}\right) \lp
\end{equation}

The excitation temperature $T_{\rm ex}$ is obtained from the $^{12}$CO
intensity. First, the maximum intensity in the spectrum of each
pixel is found. This
quantity is denoted by $T_{\rm max}$. The excitation
temperature is calculated by solving the following equation
\begin{equation}
T_{\rm max}=\frac{h\nu}{k}\left[\frac{1}{e^{\frac{h\nu}{kT_{\rm
ex}}}-1}-\frac{1}{e^{\frac{h\nu}{kT_{\rm bg}}}-1}\right] \lc
\end{equation}
where $h$, $k$ and $\nu$ are Planck's constant, Boltzmann's
constant, and the central frequency of $^{12}$CO $J=1\to 0$ line
(115.27 GHz), respectively.

The total number of H$_2$ molecules in a core as traced by $^{13}$CO
is
\begin{equation}
\Sigma_{\rm H_2}=\frac{N_{\rm tot, ^{13}CO}(D\Delta)^2}{[^{13}\rm
CO]/[\rm H_2]} \lc
\end{equation}
where the distance of the Taurus molecular cloud, $D=140\ {\rm pc}$,
the pixel size of the data cube, $\Delta=20''$, the $^{13}$CO to
H$_2$ abundance ratio $[^{13}\rm CO]/[\rm H_2]$ is taken to be
$1.7\times 10^{-6}$~\citep{CO_H_ratio_2}. Calculation of the mass of
the $^{13}$CO-traced core is then straightforward
\begin{equation}
M=\beta m_{\rm H_2}\Sigma_{\rm H_2}\lc
\end{equation}
where $\beta =1.39 $ converts the hydrogen mass to total mass taking
into account of the abundance of helium~\citep{Wilson1994}.

We then correct for the depletion of $^{13}$CO by using the recipe
of \cite{Whittet2007}, who measured the column density of CO and
CO$_2$ ices toward a sample of stars located behind the Taurus
molecular cloud and obtained the following empirical relationship
\begin{equation}
\frac{N({\rm CO})_{\rm ice}}{10^{17}({\rm cm^{-2}})}=0.4(A_V-6.7),\
\ A_V>6.7{\rm mag}\lc
\end{equation}

\begin{equation}
\frac{N({\rm CO_2})_{\rm ice}}{10^{17}({\rm
cm^{-2}})}=0.252(A_V-4.0),\ \ A_V>4.0{\rm mag}\lp
\end{equation}
The column density of the depleted CO is given by $N({\rm CO})^{\rm
total}_{\rm ice}=N({\rm CO})_{\rm ice}+N({\rm CO_2})_{\rm ice}$, the
total column density of CO is then $N({\rm CO})^{\rm total}=N({\rm
CO})^{\rm total}_{\rm ice}+N({\rm CO})_{\rm gas}$. Using the
abundance ratio of $^{13}$CO to $^{12}$CO, $f_{\rm
^{13}CO/^{12}CO}=1/70$ \citep{Stahl1992}, the depletion correction
for Taurus can be calculated. \cite{2MASS} obtained a good linear
correlation between the extinction and the total $^{13}$CO column
density (gas plus ice) using such a correction.  We first calculate
the average extinction in the $^{13}$CO core region, which
corresponds to a column density of $^{13}$CO and $^{13}$CO$_2$ ice.
Since a $^{13}$CO core occupies only a part of the velocity
channels, the ratio of the $^{13}$CO integrated intensity and total
intensity of the $^{13}$CO core region, $I_{\rm core}/I_{\rm total}$
is calculated and the mass correction for a $^{13}$CO core is
\begin{equation}
\Delta M=f_{\rm ^{13}CO/^{12}CO}\pi R^2 N({\rm CO})^{\rm total}_{\rm
ice}\lp
\end{equation}
where $R$ is the typical scale of a $^{13}$CO core defined in
equation~\ref{core_size}. This recipe achieves good depletion
correction at the arc-minute scale.

The FWHM line width $\Delta V_{\rm FWHM}$ is calculated from the
velocity dispersion of each $^{13}$CO core, $\Delta v$, which is the
standard deviation of the velocity value about centroid velocity,
weighted by the corresponding pixel data value.
\begin{equation}
{\Delta V_{\rm FWHM}}=2\sqrt{2\ln 2}\Delta v \lp
\end{equation}
The virial mass is also estimated from $\Delta v$ \citep{Pressure_confine}
\begin{equation}
M_{\rm vir}=\frac{5\Delta v^2 R}{G}\lp
\label{virial_mass}
\end{equation}
This equation describes a balance between the self gravity and
combined thermal and nonthermal motions of a $^{13}$CO core,
neglecting external pressure and the magnetic field. A core is
considered gravitationally bound when its mass exceeds $M_{\rm
vir}$.

In dealing with the extinction map, the hydrogen column density is
estimated from optical extinction \citep{extinction_column}
\begin{equation}
\frac{N_{\rm H}}{\rm cm^{-2}}=2.2\times 10^{21} \left(\frac{A_V}{\rm
mag}\right)\lp
\end{equation}

\begin{figure}[htb]
\includegraphics[width=18cm]{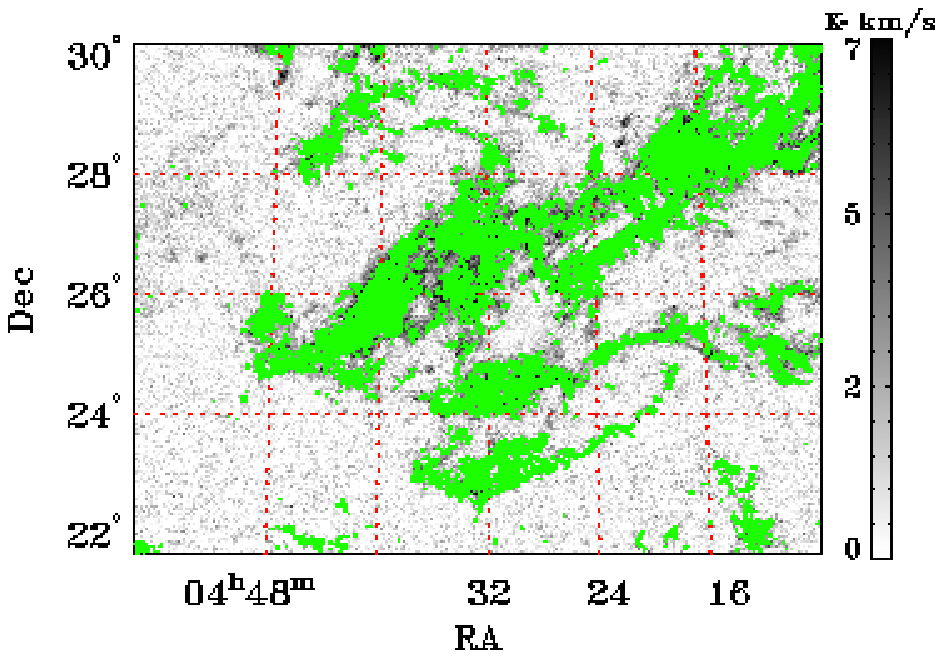}
\caption{ The cores found with CLUMPFIND routine are overlayed on
the \13co\ total intensity map. It is clear from this figure that
the cores are "crowded". In this case, CLUMPFIND cannot extract
cores properly.
 \label{overlay_cl}}
\end{figure}

We have applied GAUSSCLUMPS with the fitting parameters listed in
table~\ref{gaussclumpspara} to the Taurus region to extract cores.
We have also applied CLUMPFIND with the fitting parameters  listed
in table~\ref{clumpfindpara}. In both cases, a background was first
subtracted with the FINDBACK procedure using the same parameters
listed in the tables. The cores fitted by GAUSSCLUMPS are shown in
Figure~\ref{fig1} and those by CLUMPFIND in Figure~\ref{overlay_cl}.
There are many more smaller cores found by CLUMPFIND, especially in
the relatively diffuse region. This is attributable to CLUMPFIND's
requirement to assign one pixel to one particular core without
'splitting' a pixel in possible overlapping configurations. We rely
on the cores found by GAUSSCLUMPS in the subsequent analysis, since
CLUMPFIND cannot split overlapping cores properly. In a study of
core extraction with CLUMPFIND, \cite{Pineda2009} suggest not to
blindly use CLUMPFIND to derive mass functions in any "crowded"
situation.

Parameters of 3D \13co\ cores and those cores found in the smoothed
\13co\ data cube can be found in tables~\ref{tab:clumps}
and~\ref{tab:clumps_smooth}, respectively. In table
\ref{tab:clumpsEx}, we give parameters of the cores derived from the
2MASS extinction map.

\section{Results}
\label{sec:results}

\begin{figure}[htb]
\includegraphics[width=15cm]{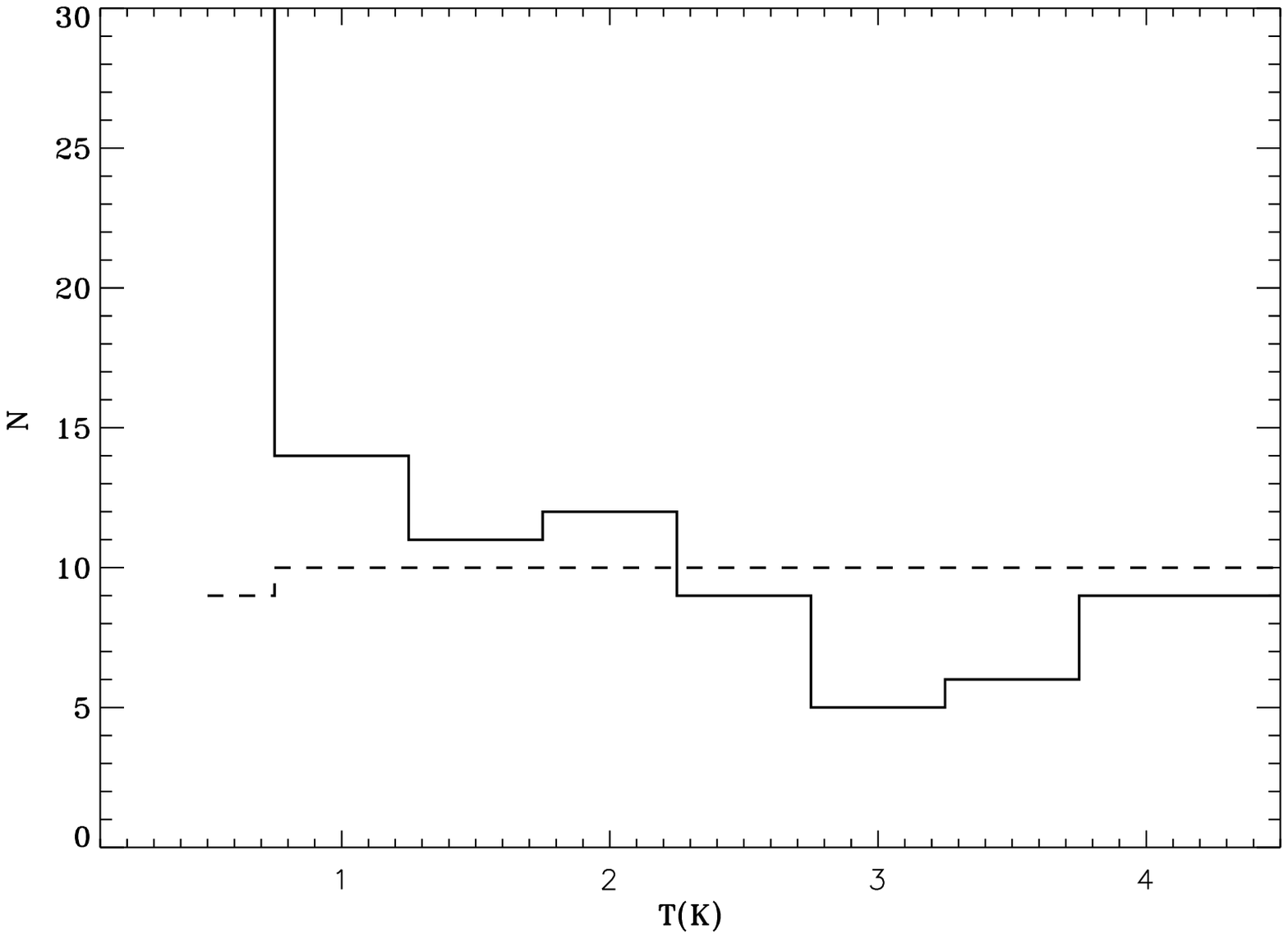}
\caption{Core extraction from simulated data for $^{13}$CO by using
GAUSSCLUMPS. The dashed line shows the distribution as a function of
temperature of the simulated cores, while the solid line describes
the cores extracted. When the core peak intensity is higher than
$7\sigma \sim 0.7 $ K, the difference is small. For peak intensity
lower than $7\sigma$, GAUSSCLUMPS tends to extract more cores than
are actually present. In this work only $^{13}$CO cores with peak
intensity higher than 7$\sigma$ are included.
 \label{simulate_data}}
\end{figure}

We have performed a test on GAUSSCLUMPS by fitting to simulated
data. We put 100 Gaussian components with random sizes and random
locations into an empty simulation cube (an empty 3D array) with the
size of region 1 in Figure \ref{fig1}. White noise (with $\sigma\sim
0.1$ K, a typical value across the Taurus region.) is also added to
the whole cube as a background. This simulated data cube is then
fitted by using the GAUSSCLUMPS tool. As a result, we find that for
cores with peak intensity higher than 0.7K ($7\sigma$), the
distribution of the cores found by GAUSSCLUMPS is similar to the
simulated distribution. Figure \ref{simulate_data} gives numerical
results. For lower masses, GAUSSCLUMPS tends to find more cores
\citep{Stutzki1990}. Similar simulations have been done for the 2D
data ($^{13}$CO total intensity data and extinction data), where a
threshold of $5\sigma$ is found to ensure agreement between measured
and simulated distributions.

From the GAUSSCLUMPS fitting, we select cores with peak intensity
higher than $7\sigma\sim 0.7$ K for $^{13}$CO data cube based on the
simulation mentioned above, $5\sigma\sim 1.3$ K$\cdot$km/s for
$^{13}$CO total intensity, and $5\sigma\sim 1.5$ mag for the
extinction map, where $\sigma$ is the variance of data (the RMS).
Other fitting parameters can be found in
tables~\ref{gaussclumpspara}, \ref{clumpfindpara}, and
\ref{extinctionpara}.

Some filamentary structures (ellipses with a large axis ratio) were
found in both the extinction map and the $^{13}$CO total intensity
map. They comprise only a small fraction of the total mass (less
than 10\%). They do not show up in the fitting to the $^{13}$CO data
cube, which means they are not "coherent", meaning that, the
velocity variation within the structure cannot be described by a
Gaussian in velocity space. Since the stability of a filament is
different from that of a core \citep{Lombardi2001}, we filter out
those gaussian components that have an large axis ratio (major/minor
$>$10) in the analysis. We thus obtain a sample of 765 $^{13}$CO
cores. This is the largest sample of spectral line defined cores for
a contiguous region
\citep[cf.][]{Tatematsu1993,Aso2000,Ikeda2007,Tatematsu2008,Ikeda2009,Ikeda2009_2,Ikeda2011}.

\begin{figure}[htb]
\begin{tabular}{c}
\includegraphics[width=18cm]{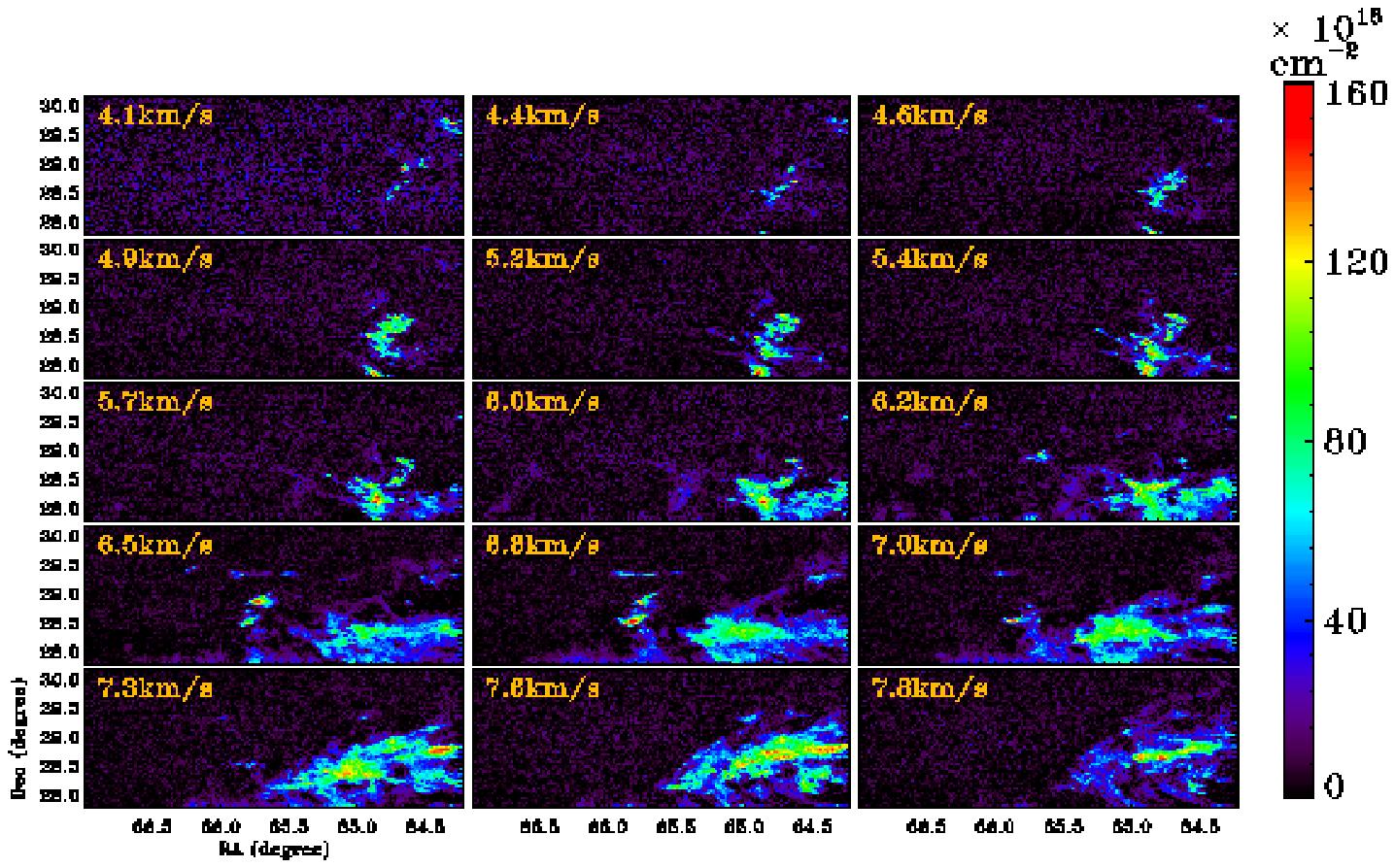}\\
\end{tabular}
\caption{$^{13}$CO channel maps of the region 11 in
figure~\ref{fig1}. The intensity changes drastically between
channels, which is in stark contrast with the region shown in
figure~\ref{channel2} and suggests velocity incoherence in region
11. This is the reason why no $^{13}$CO core can be fitted by a
three dimensional (x,y,v) gaussian in this region. \label{channel}}
\end{figure}

\begin{figure}[htbp]
\begin{tabular}{c}
\includegraphics[width=18cm]{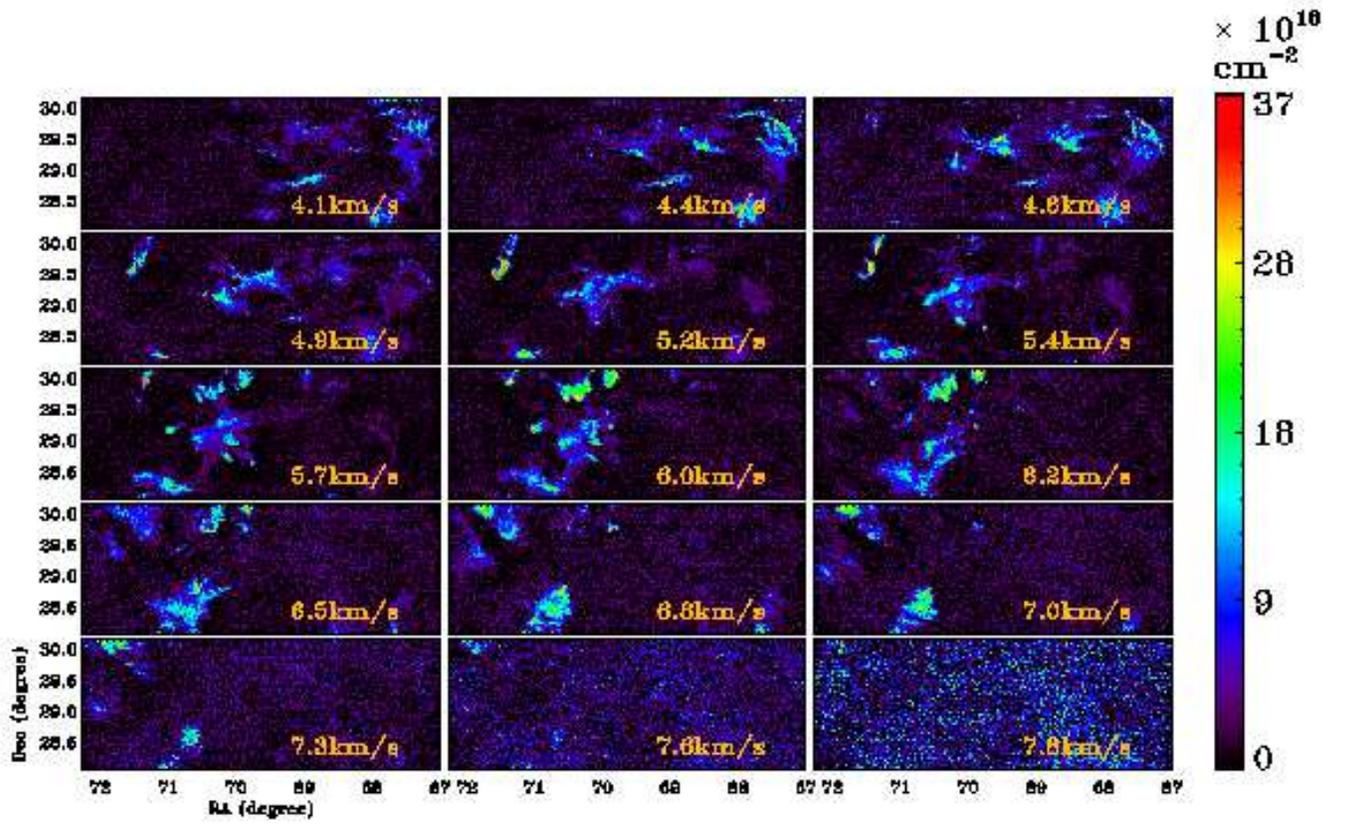}\\
\end{tabular}
\caption{Channel maps of region 5 in figure~\ref{fig1}. The
intensity here also changes drastically between channels. Only one
$^{13}$CO core is found in this region. \label{channel3}}
\end{figure}

\begin{figure}[htbp]
\begin{tabular}{c}
\includegraphics[width=18cm]{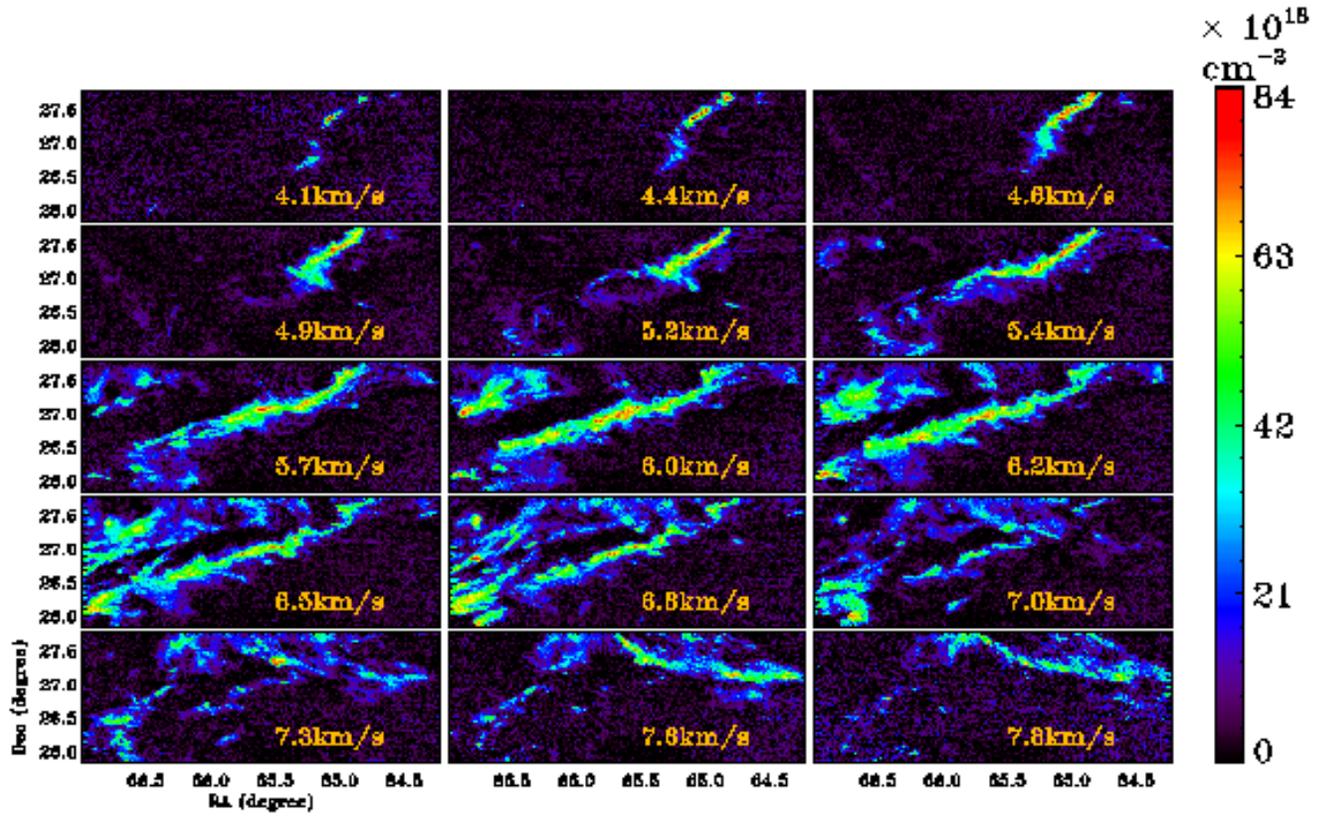}\\
\end{tabular}
\caption{Channel maps of region 10 in figure~\ref{fig1}. The main
features are coherent through multiple channels and in consequence
many $^{13}$CO cores are found in this region. \label{channel2}}
\end{figure}

Figure~\ref{fig1} shows the $^{13}$CO integrated intensity map
overlayed with the cores obtained by Gaussian fitting to the
$^{13}$CO data cube. Most of the regions with \13co\ emission are
found to contain cores. However, despite the strong \13co\ emission
in region 11 and 5, no $^{13}$CO core is found in the former and
only one $^{13}$CO core is found in the latter. This is due to the
rapid variation of intensity between velocity channels, which is
clear in figures~\ref{channel} and~\ref{channel3}, which are the
channel maps of region 11 and region 5, respectively. Gas in region
11 is likely to be affected by the young protostellar cluster within
it~\citep{L1495_B10}. In contrast to these two regions, the \13co\
emission in all other regions containing $^{13}$CO cores is found to
change gradually in velocity (see, e.g.\ figure~\ref{channel2}).

\clearpage

\begin{figure}[htbp]
\begin{tabular}{cc}
\includegraphics[width=8cm]{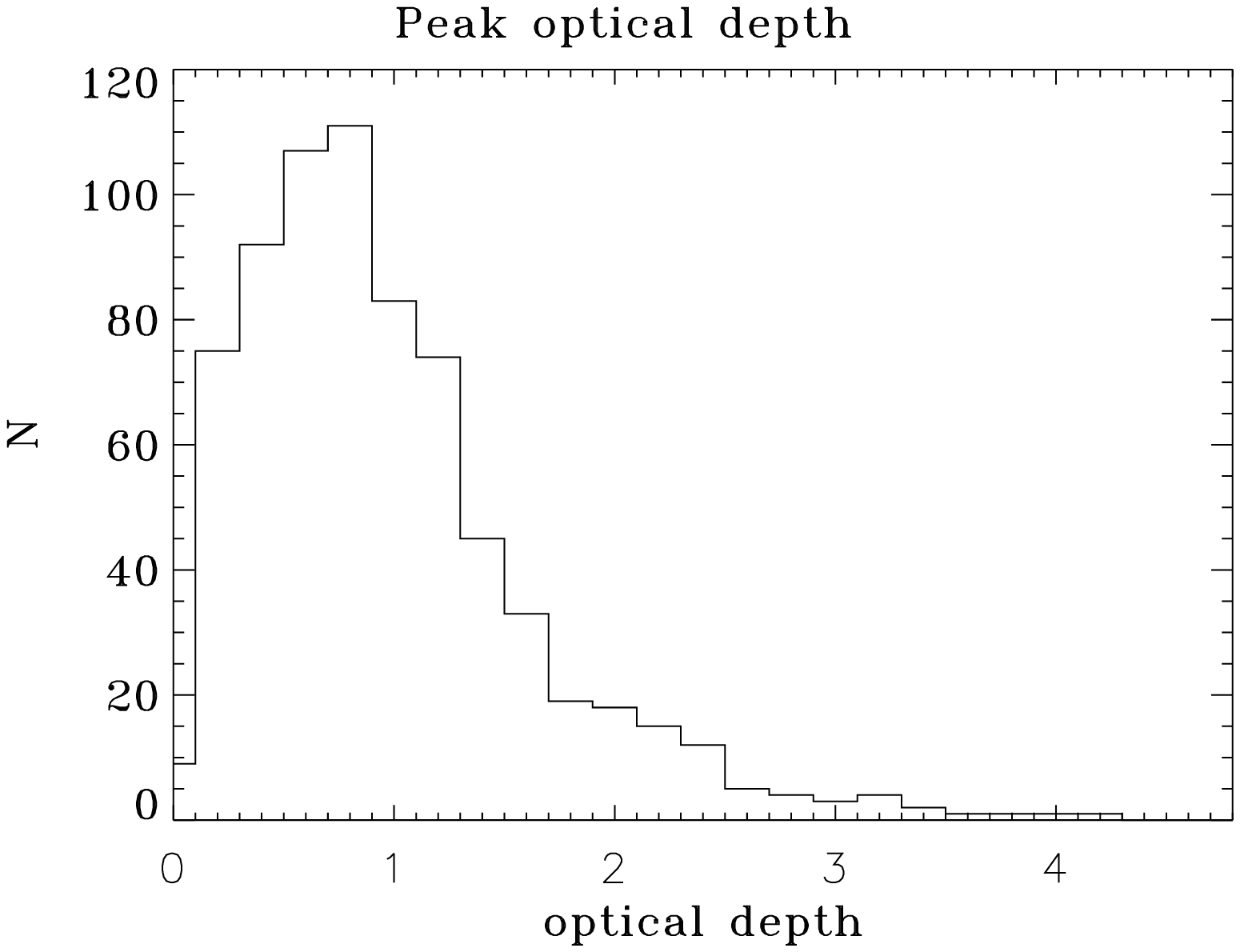}&\includegraphics[width=8cm]{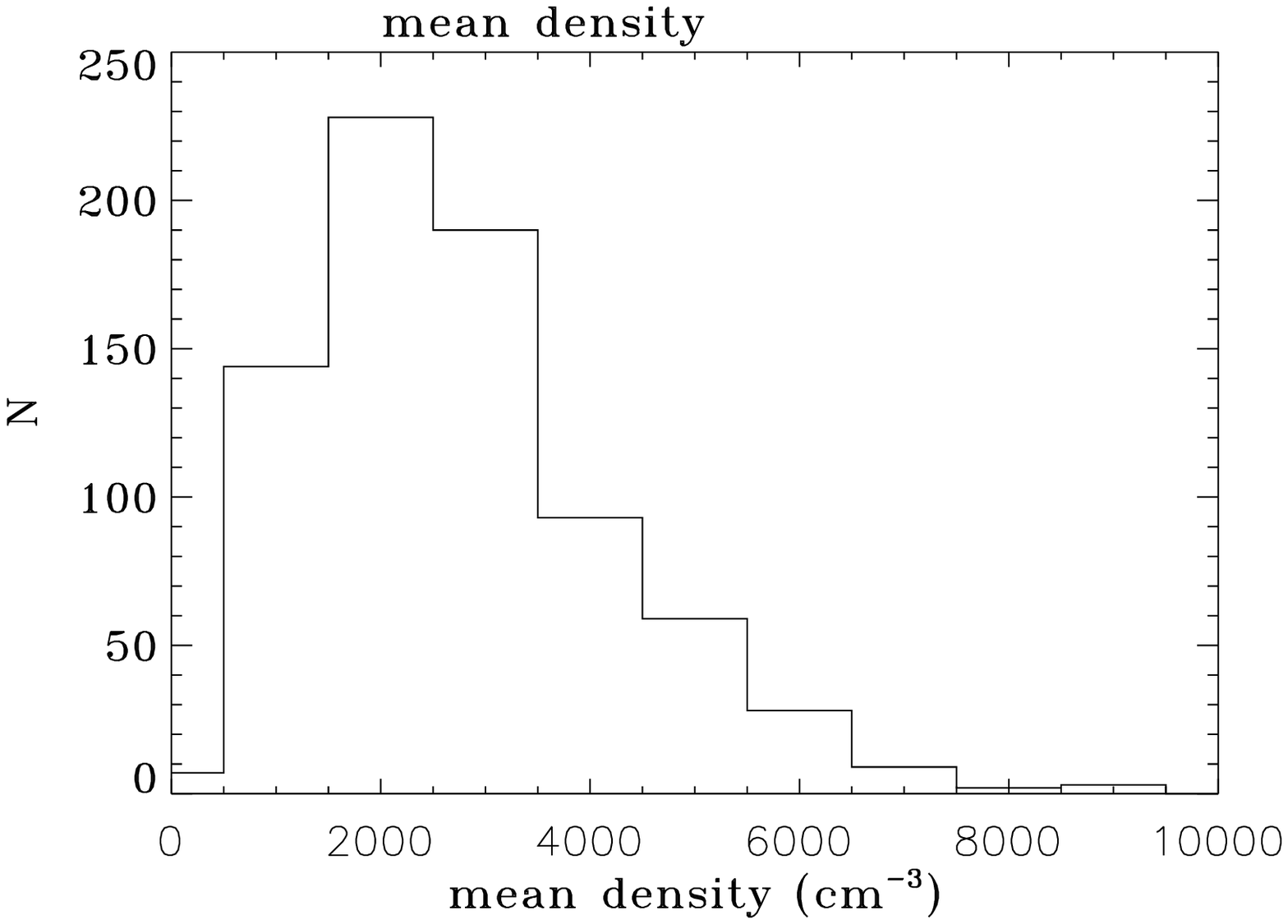}\\
\end{tabular}
\caption{The distribution of the peak optical depth (left panel) and
mean $\rm H_2$ density (right panel) of the $^{13}$CO cores.
\label{depth_density}}
\end{figure}

The distributions of the peak optical depth and the mean $\rm H_2$
density of the $^{13}$CO cores are shown in
Figure~\ref{depth_density}. The mean $\rm H_2$ density is calculated
by dividing the total number of $\rm H_2$ in a $^{13}$CO core by its
volume, assuming its size along the line of sight is the same as its
typical size (geometrical mean of major and minor, 2 times the
typical radius in equation~\ref{core_size}). Most of the $^{13}$CO
cores have a typical $J=1\to 0$ optical depth of $\sim 0.8$ and a
typical mean $\rm H_2$ density of $\sim 2000\ \rm cm^{-3}$. The
recipes that we have adopted for optical depth correction and
excitation state correction are adequate for cores in Taurus.

By looking at the motion of the $^{13}$CO cores, we find a group of
$^{13}$CO cores with systematically different line of sight
velocities (see the lower corner of figure~\ref{overlay_velocity}),
which suggests these $^{13}$CO cores may not belong to the main
cloud. After excluding these, we are left with 588 $^{13}$CO cores.

\begin{figure}[htb]
\centering
\includegraphics[width=18cm]{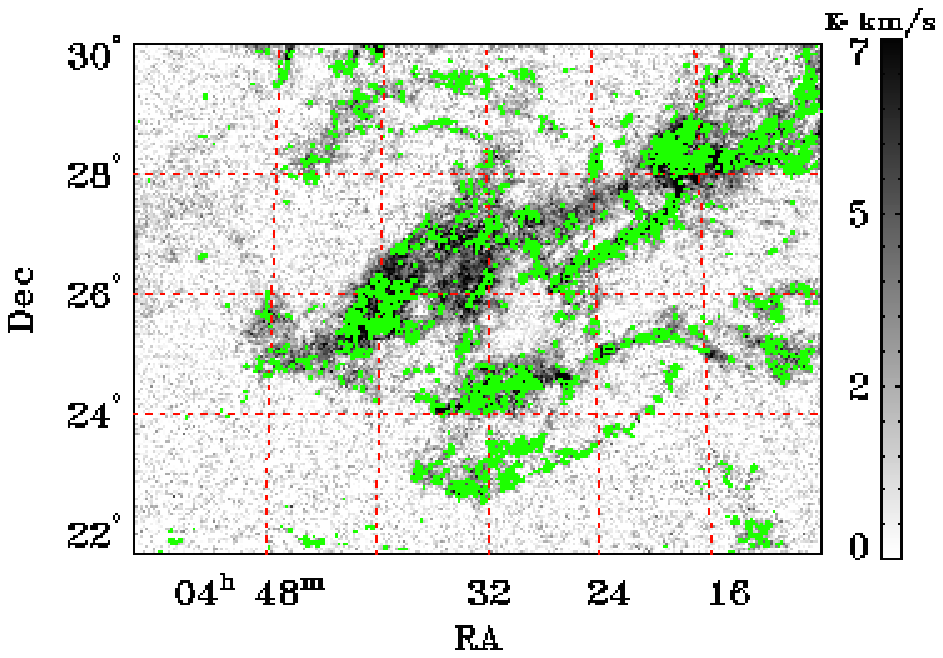}
\caption{ The cores found through a 2D Gaussian fitting of the \13co total intensity map are overlaid on the $^{13}$CO total intensity map of the whole Taurus region.
The red dashed lines are the coordinate lines of constant right ascension and declination.\label{overlay2}}
\end{figure}

For comparison, we also use GAUSSCLUMPS to fit cores in the
$^{13}$CO integrated intensity map with the procedures used for core
fitting in the $^{13}$CO data cube. Figure~\ref{overlay2} shows the
$^{13}$CO integrated intensity map overlaid with cores obtained by
Gaussian fit of the $^{13}$CO integrated intensity map, which is a
2D core fitting. Compared with the 3D core fitting to the \13co\
data cube, there are cores found in the 2D core fitting within the
regions having relatively abrupt, significant velocity variations
seen in the \13co\ data. This indicates that velocity is essential
for reasonably defining cores in a complex cloud such as Taurus. The
physical plausibility of requiring that a core have a relatively
well-defined velocity in additional to a compact spatial structure
eliminates many spurious cores that are identified in the 2D data
alone.

\begin{figure}[htb]
\centering
\includegraphics[width=18cm]{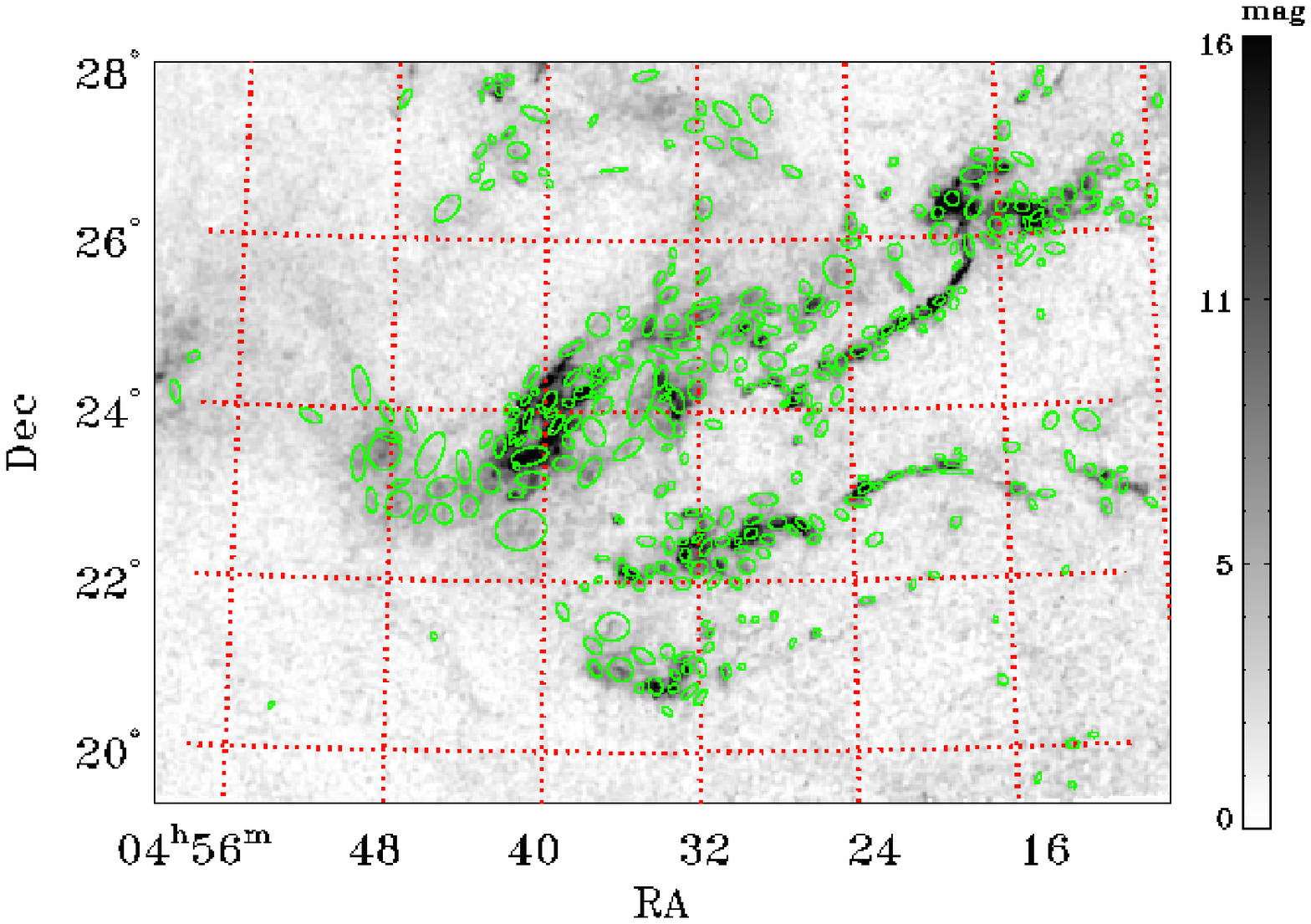}
\caption{ The cores found through a 2D Gaussian fitting of the
extinction map are overlaid on the dust extinction map of the whole
Taurus region. They are generally larger than the \13co\ cores. The
red dashed lines are the coordinate lines of constant right
ascension and declination. \label{overlayEx}}
\end{figure}

Figure~\ref{overlayEx} shows dust extinction map overlayed with
cores obtained by Gaussian fit to the 2MASS extinction map. The
200\arcsec\ spatial resolution of 2MASS extinction map is about 5
times coarser than that of \13co\ map. The extinction cores are
generally larger than those found in \13co\ data. We perform an
experiment by smoothing the \13co\ data to the same spatial
resolution as 2MASS extinction and perform the core fitting. The
fitted cores are also generally larger than those found in the
original \13co\ data, confirming that the cores found in the 2MASS
data are generally unresolved even at the modest distance of Taurus
(see table~\ref{tab:clumps_smooth}).

\subsection{Mass distribution of the cores}

We analyze the CMF in two functional forms, power law
(equation~\ref{power_law}) and log-normal
(equation~\ref{lognormal}). Direct fitting of power laws to a
cumulative mass function can be erroneous due to the natural
curvature in the cumulative CMF (see the appendix
in~\cite{core_mass_function}). Therefore, we adopt a Monte Carlo
approach similar to that used in ~\cite{core_mass_function}. We
first generate a random sample of cores with one of the above
distributions. For a power law distribution, there are three
parameters, $M_{\rm max}$, $M_{\rm min}$, and $\gamma$, while for a
log-normal distribution, there are two, $\sigma$ and $\mu$. The
cumulative distribution $C(M_i),\ i=1,...,n$ of the random sample is
then fitted to that of the sample found in the data $C_0(M_i),\
i=1,...,n$ by minimizing a $\chi^2$ function defined by
\begin{equation}
\chi^2\equiv \sum_{i=1}^{n}\frac{[C(M_i)-C_0(M_i)]^2}{C_0(M_i)^2}\lp
\end{equation}

\begin{figure}[htb]
\begin{tabular}{cc}
\includegraphics[width=8cm]{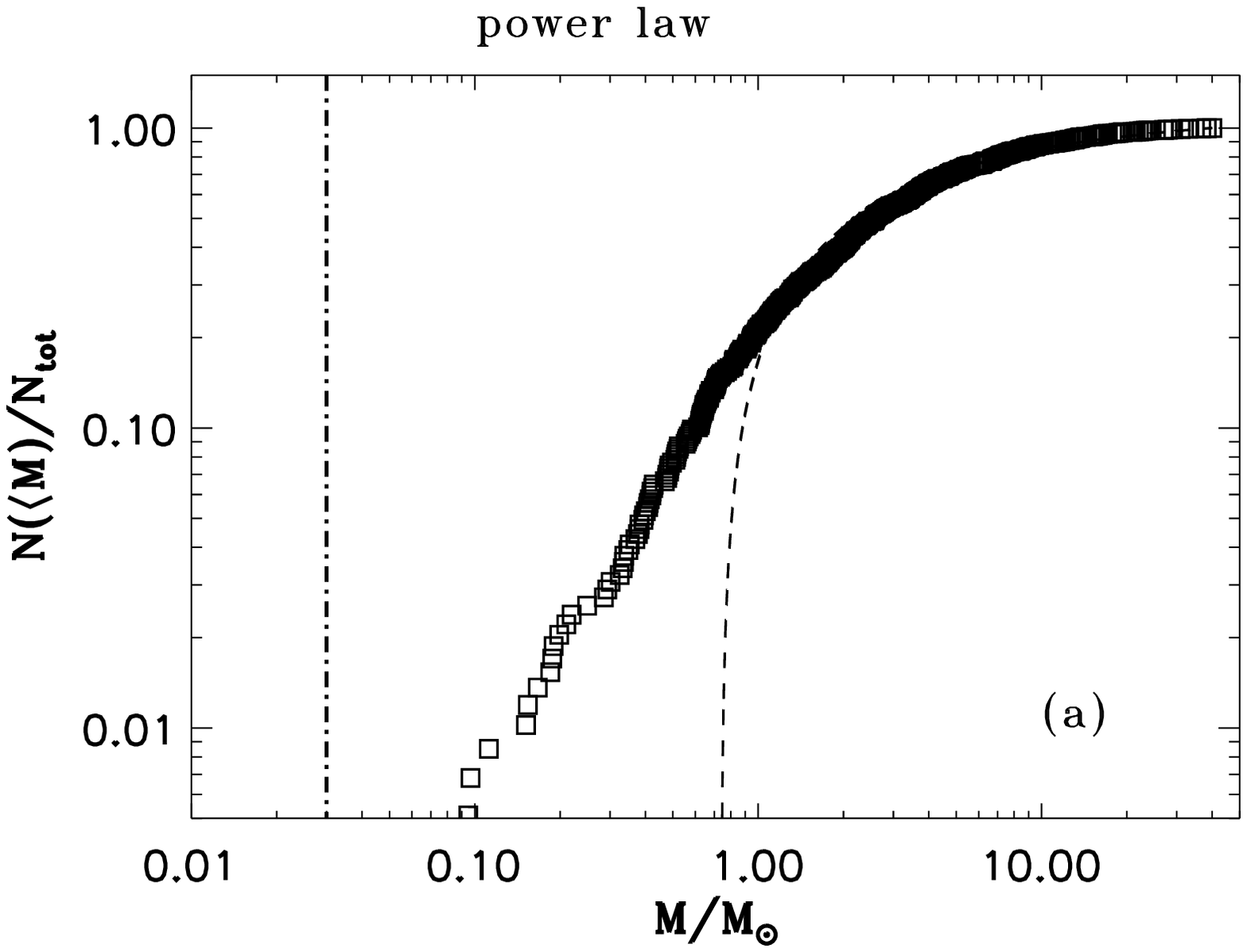} & \includegraphics[width=8cm]{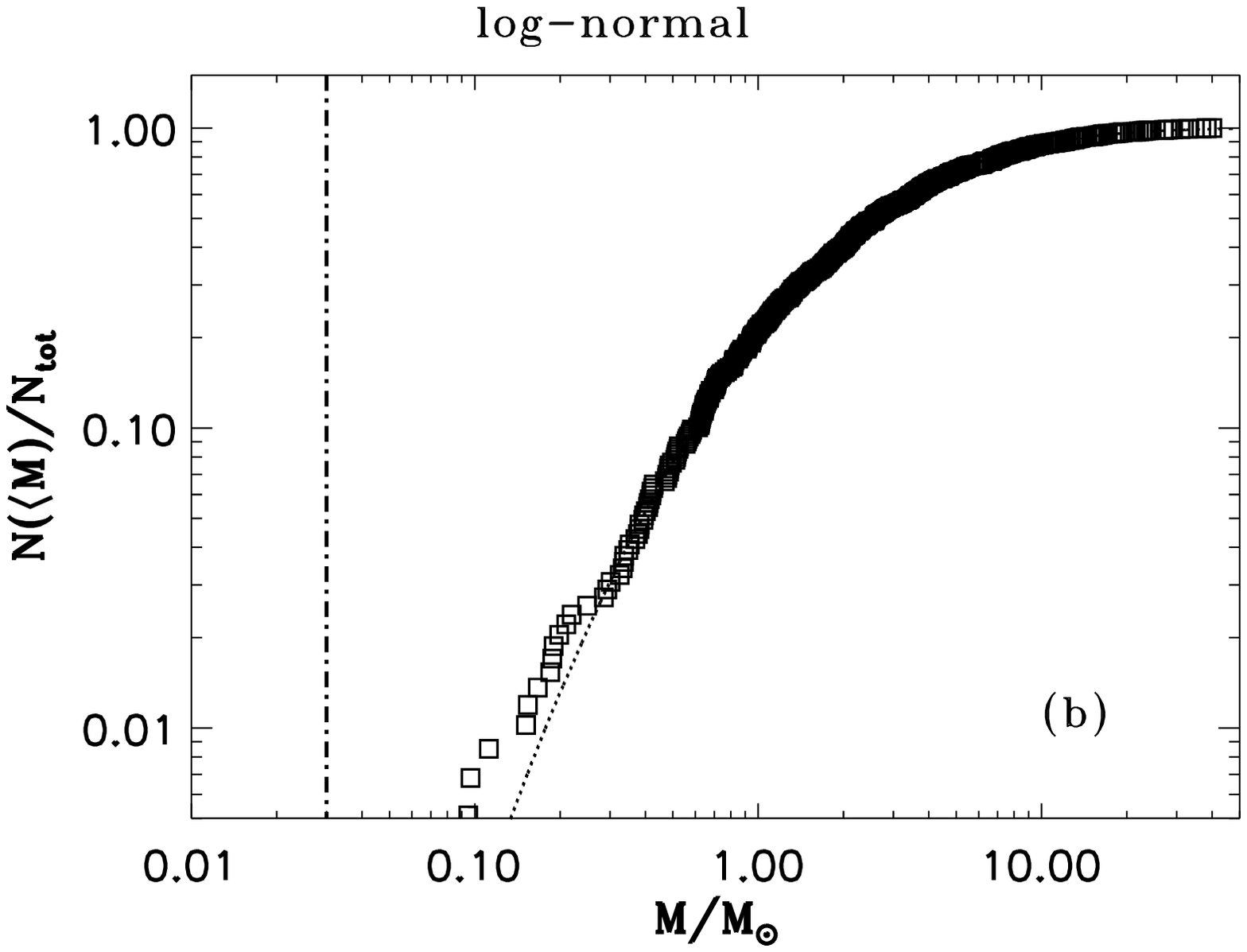}\\
\end{tabular}
\caption{ The mass distribution of the $^{13}$CO cores found by
GAUSSCLUMP fitting to the $^{13}$CO data cube. The left panel shows
the best fit power law distribution, while the right panel shows the
best fit log-normal distribution, with the latter giving a better
fit over the full mass range. The minimum detection mass for the
$^{13}$CO cube is $0.03\ M_{\odot}$, indicated by the vertical
dash-dotted line. We have fitted the mass function to $^{13}$CO
cores more massive than the minimum detection mass. The fitting
results are : (a) power-law distribution with $dN/d\log M\propto
M^{-0.53}$, $M_{\rm max}=40.03\ M_{\odot}$, $M_{\rm min}=0.74\
M_{\odot}$. (b) log-normal distribution with $\mu=0.95$,
$\sigma=1.15$. \label{massfunction}}
\end{figure}

\begin{figure}[htb]
\begin{tabular}{cc}
\includegraphics[width=8cm]{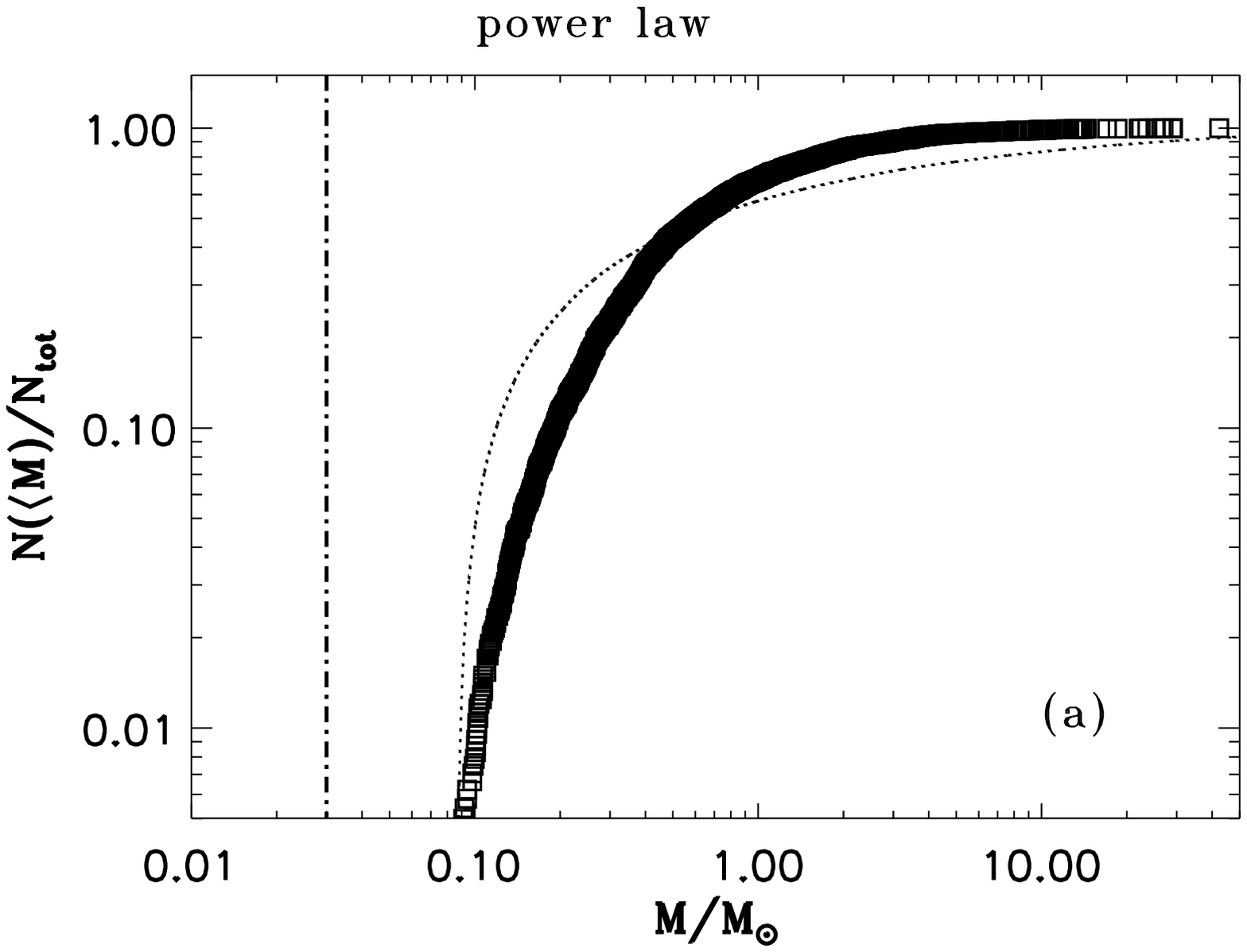} & \includegraphics[width=8cm]{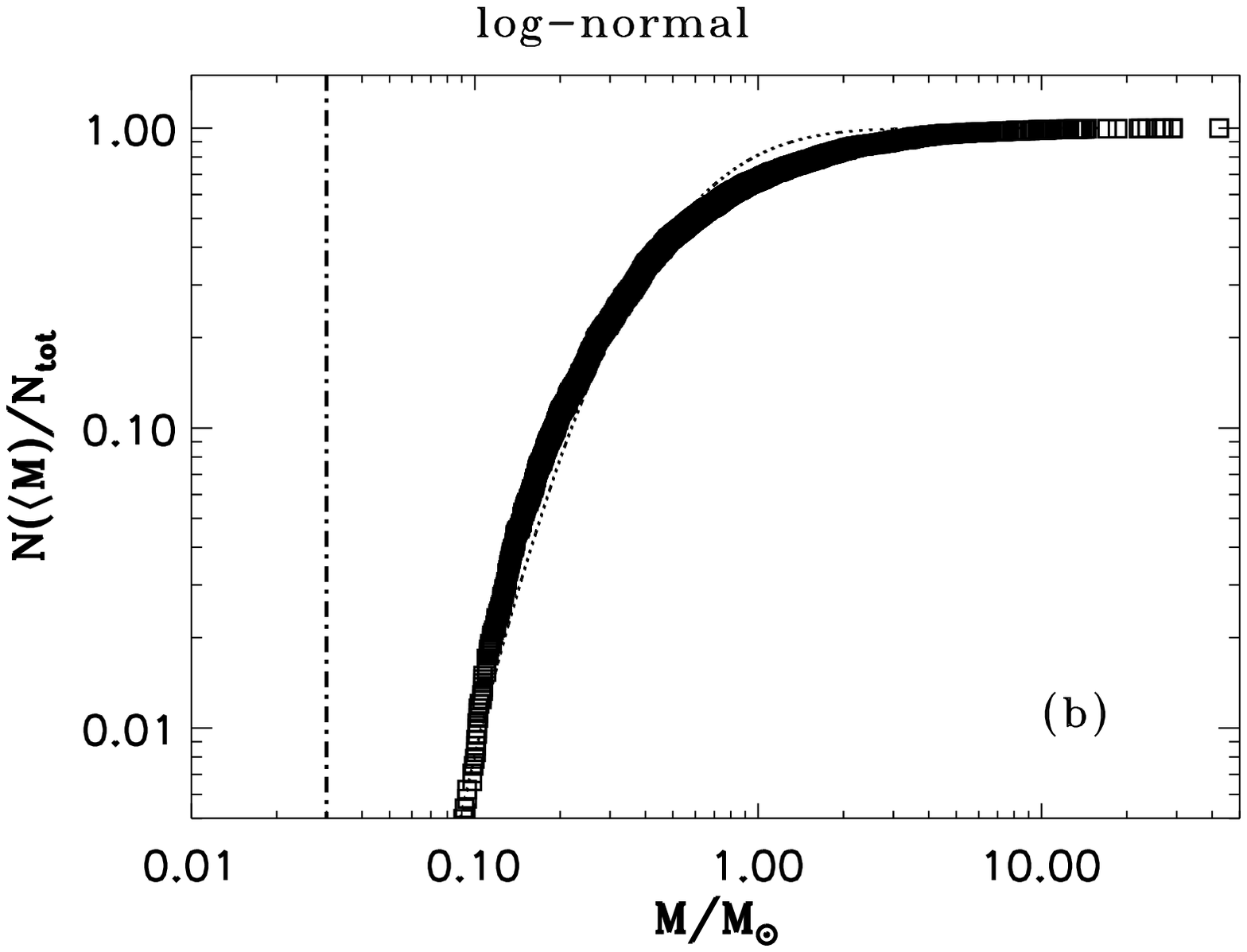}\\
\includegraphics[width=8cm]{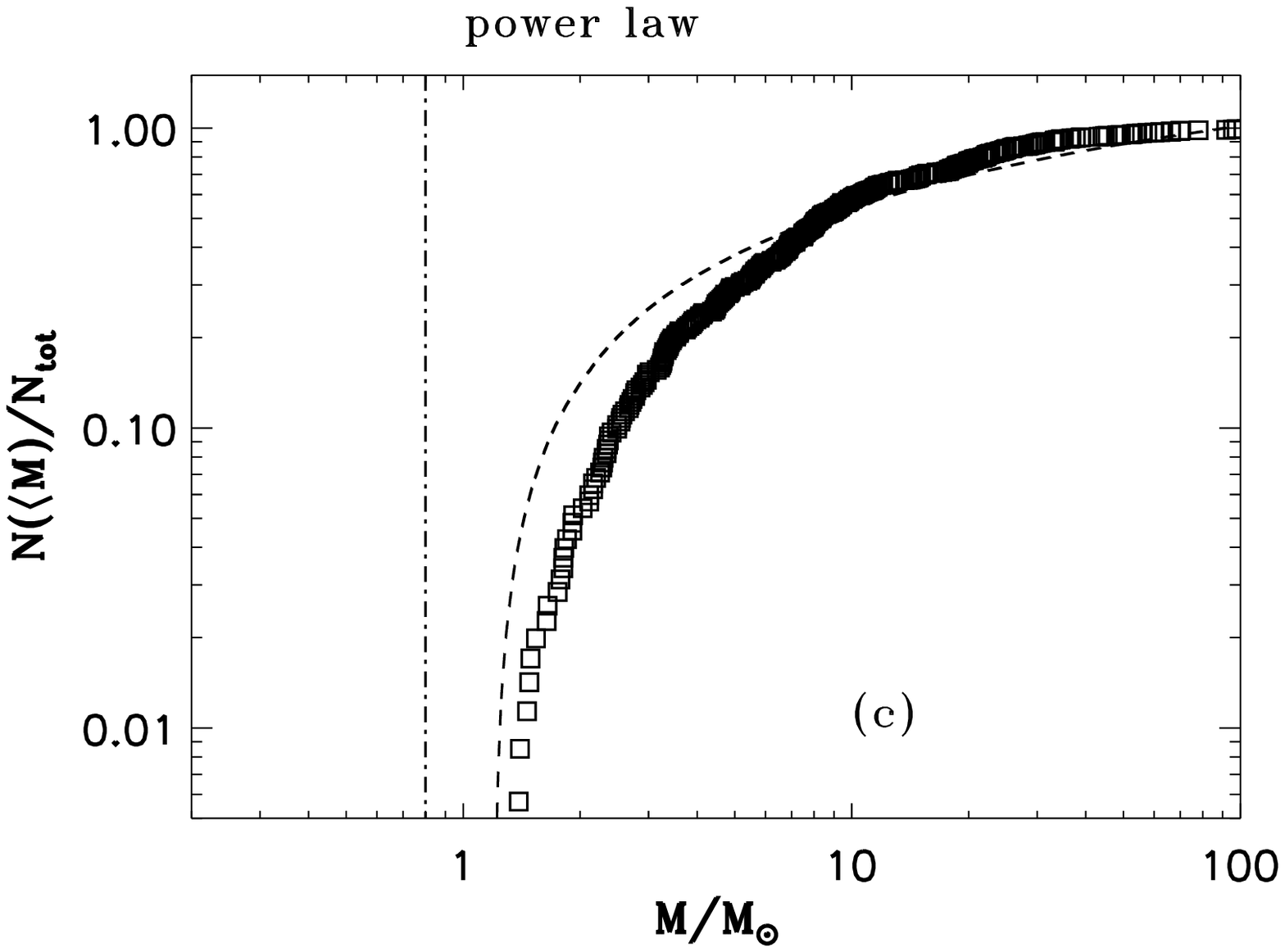} & \includegraphics[width=8cm]{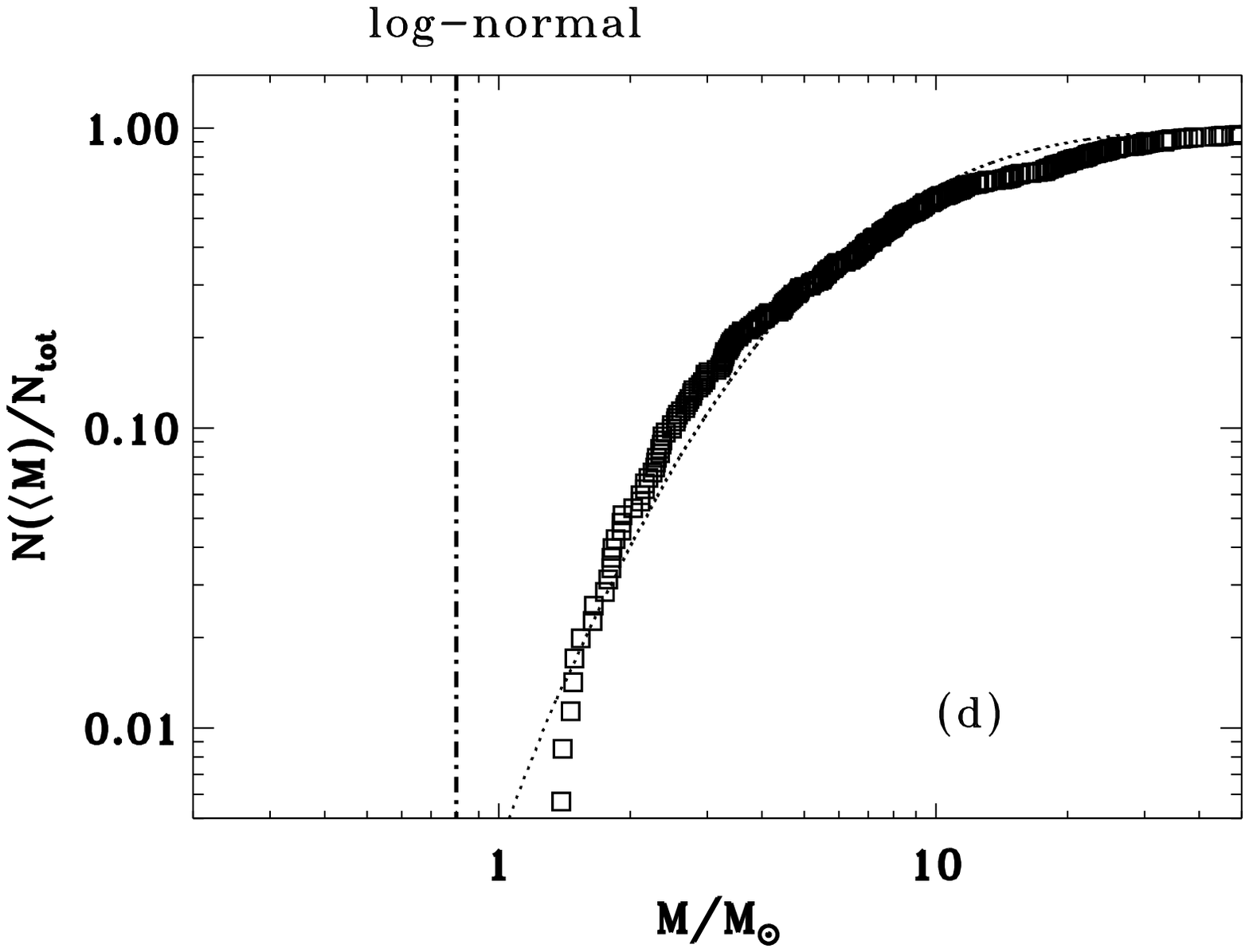}\\
\end{tabular}
\caption{ The mass distribution of the cores found by GAUSSCLUMP
fitting to the $^{13}$CO total intensity map ({\it upper row}) and
to the extinction map ({\it lower row}). The panels on the left are
fitted with power law distributions, while those on the right are
fitted with log-normal distributions. The minimum detection masses
are $0.03\ M_{\odot}$ and $0.8\ M_{\odot}$ for the $^{13}$CO data
cube and the extinction map, respectively, and are indicated by
vertical dash-dotted lines. The fitted results are: (a) $dN/d\log
M\propto M^{-0.30}$, $M_{\rm max}=106.48\ M_{\odot}$, $M_{\rm
min}=0.13\ M_{\odot}$. (b) log-normal distribution with $\mu=-0.62$,
$\sigma=0.70$. (c) $dN/d\log M\propto M^{-0.1}$, $M_{\rm max}=92.0\
M_{\odot}$, $M_{\rm min}=1.2\ M_{\odot}$. (d) log-normal
distribution with $\mu=2.04$, $\sigma=0.77$. The cores from
$^{13}$CO total intensity and the extinction cores are both fitted
better with a log-normal distribution than with a power law.
\label{massfunction2}}
\end{figure}

The resulting mass function of the $^{13}$CO cores can be seen in
figure~\ref{massfunction}. This mass distribution can be fitted much
better by a log-normal function than by a power law, which only fits
the higher mass end. The fitting results are similar for cores based
on 2D fitting.  The mass distribution that is based on 2D fitting of
the $^{13}$CO integrated intensity can be fitted better with a
lognormal function. The mass function of 2MASS extinction cores can
also be fitted better with a log-normal function than with a power
law (figure~\ref{massfunction2}).
Thus log-normal is a better representation of the mass
distribution of cores than power law in the case of cores
represented by $^{13}$CO, $^{13}$CO total intensity, and 2MASS
extinction.

To examine the reliability and the completeness of the cores found, we calculate the minimum detectable mass for a core of certain size \citep{core_mass_function}
\begin{equation}
M=M_{\rm point}\times\sqrt{N}\times\sqrt{M},
\end{equation}
where $M_{\rm point}$ is the minimum detectable mass of a point
source, and $N$ and $M$ are the number of pixels and number of
velocity channels occupied by a core, respectively. We found this
minimum detection mass is about 0.03 $M_{\odot}$ for a core as large
as the smoothing scale. There is also another way to estimate the
completeness~\citep{Pineda2009}, with which similar result is
obtained. All the cores found in the $^{13}$CO data cube are larger
than this minimum mass. For the extinction map, the minimum
detectable mass of a core is found to be about 0.8 $M_{\odot}$.

We used bootstrap \citep[see e.g.][]{Press1992} to estimate the
uncertainty of the fitting. The resulting uncertainty is extremely
small when the sample size is large. This means that the statistical
uncertainty in our Monte-Carlo type fitting procedures is
negligible. The difference between the fitted function and the data
has to be systematic. For this reason, we do not show the numerical
values of the uncertainties.

\subsection{Energy State of $^{13}$CO Cores}
We study the energy state of $^{13}$CO cores by analyzing the mass
and line width together, which can only be accomplished through
spectral maps. The properties of the 10 most massive $^{13}$CO cores
are listed in table \ref{tab:clumps} for the $^{13}$CO data cube.

\begin{figure}[htb]
\centering
\begin{tabular}{c}
\includegraphics[width=12cm]{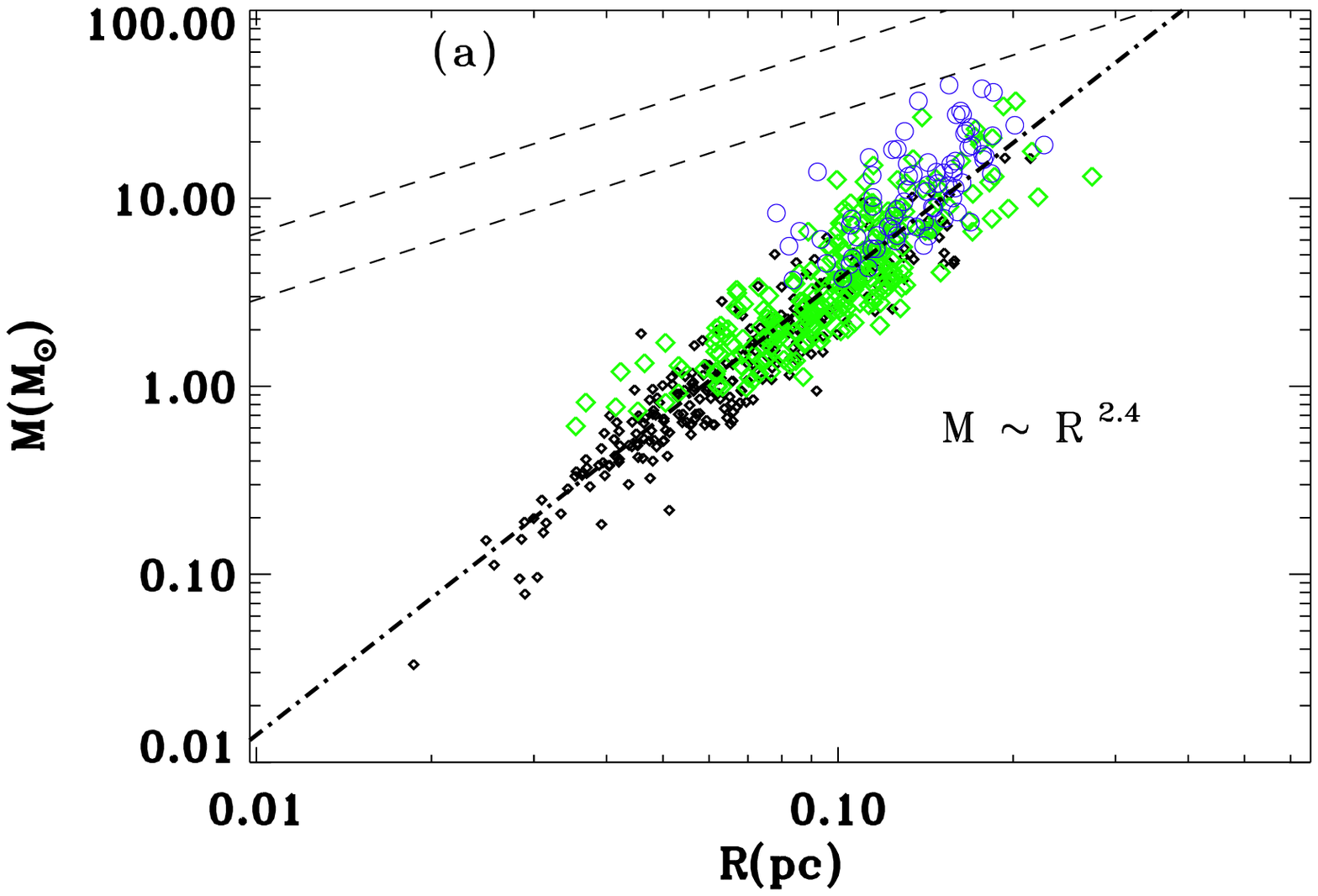}\\
\end{tabular}
\caption{ The mass-size relation of the $^{13}$CO cores found by
GAUSSCLUMPS in the $^{13}$CO data cube. The Bonnor-Ebert mass is
indicated by the dashed lines. The upper line corresponds to FWHM
linewidth of 1.5 km/s and the lower line corresponds to 1.0 km/s.The
mass-size relation is best fitted as $M\propto R^{2.6}$ indicated by
the dash-dot line. $^{13}$CO cores with virial mass to mass ratio
$M_{\rm vir}/M<2$ are denoted by blue circles; those with $2\le
M_{\rm vir}/M<5$ are denoted by green diamonds; other $^{13}$CO
cores with $M_{\rm vir}/M>5$ are denoted by black
diamonds.\label{mass_radius}}
\end{figure}

\begin{figure}[htb]
\centering
\begin{tabular}{c}
\includegraphics[width=12cm]{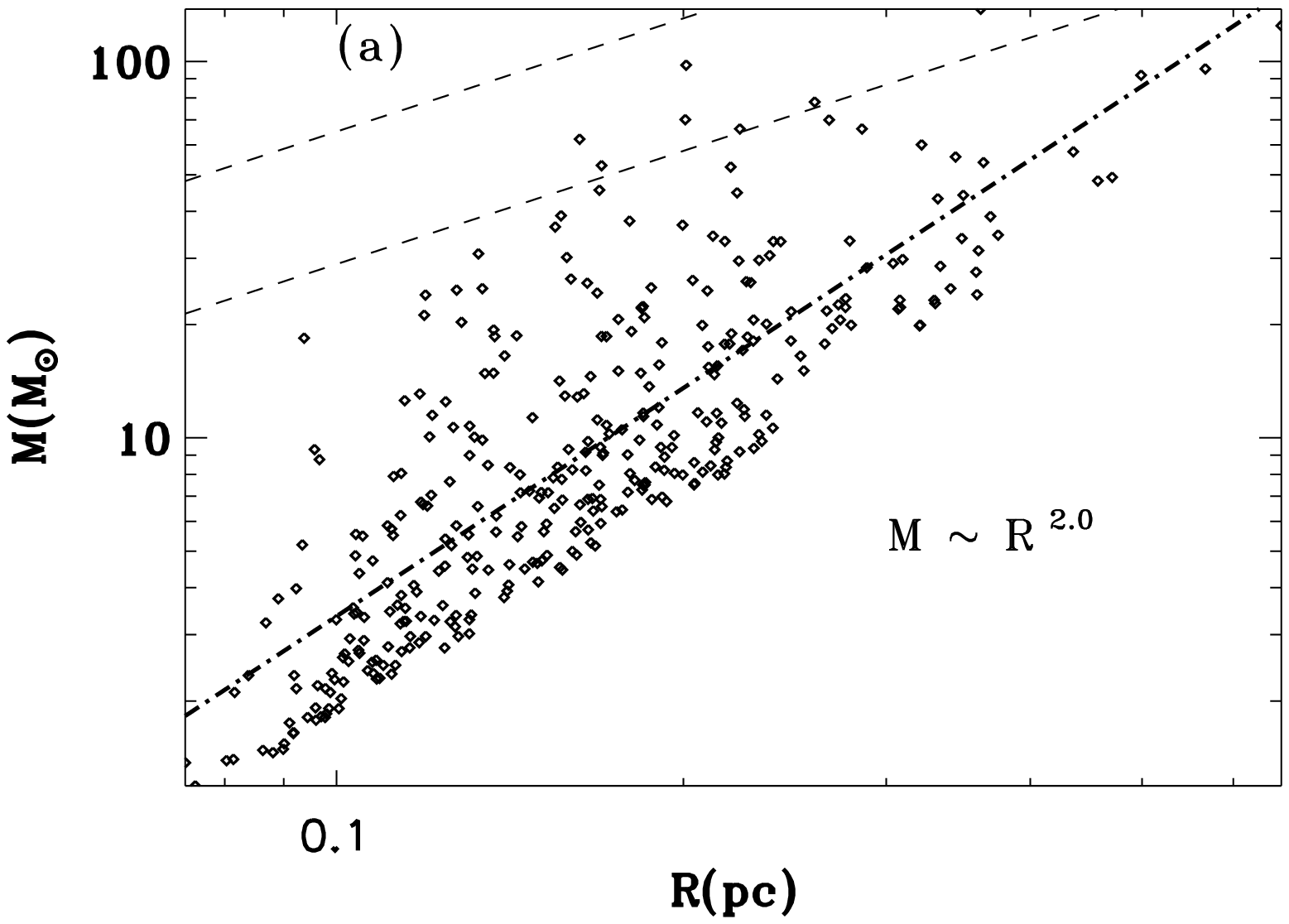}\\
\includegraphics[width=12cm]{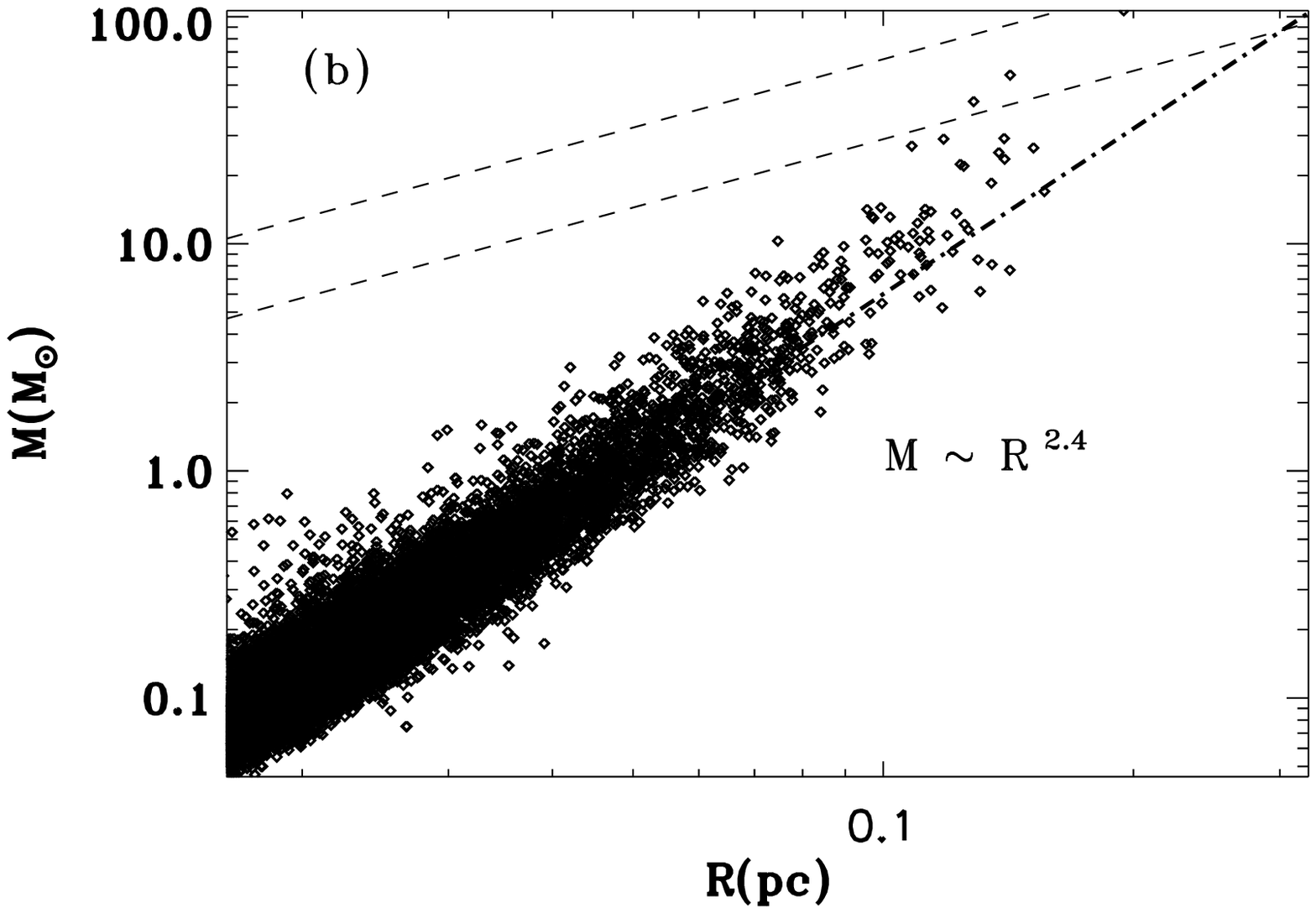}\\
\end{tabular}
\caption{ The relation between the mass and size of the $^{13}$CO
cores found in the extinction map (a) and in the $^{13}$CO total
intensity map (b) with GAUSSCLUMPS. The mass-size relationship of
extinction cores is best fitted with a power law $M\propto R^{2.0}$,
while $^{13}$CO total intensity cores can be fitted with a power law
$M\propto R^{2.4}$. The Bonnor-Ebert mass is indicated by the dashed
lines. The upper line corresponds to FWHM linewidth of 1.5 km/s and
the lower line corresponds to 1.0 km/s. \label{mass_radius2}}
\end{figure}

We investigate the mass-radius relation of the $^{13}$CO cores,
which can be seen in figures~\ref{mass_radius} and
~\ref{mass_radius2}. The core stability is studied by calculating
the Bonnor-Ebert mass-radius relation \citep{core_mass_function}, in
which the Bonnor-Ebert mass is obtained from integration of the
density profile of the Bonnor-Ebert sphere
\begin{equation}
\rho(\xi)=\frac{1}{1+(\xi/2.25)^{2.5}}\lc
\label{beprofile}
\end{equation}
where the dimensionless radius $\xi$ is defined as
\begin{equation}
\xi=r\sqrt{4\pi G\rho_c/v_s^2} \lc
\end{equation}
where $\rho_c$ is the central density and $v_s=\sqrt{kT_{\rm eq}/\mu
m_{\rm H}}$ is the effective sound speed. The equivalent temperature
$T_{\rm eq}$ is related to the FWHM line width $\Delta V_{\rm FWHM}$
with following equation ~\citep{core_mass_function},
\begin{equation}
T_{\rm eq}\equiv \frac{m_{\rm H}(3\Delta V_{\rm FWHM}^2)}{8\ln (2)
k}\lp
\end{equation}
In our sample, the FWHM line width of the $^{13}$CO cores ranges
from 0.5 km/s to 1.7 km/s. When $\xi$ is larger than $\xi_{\rm
max}=6.5$, there is no longer any stable solution. With $\xi_{\rm
max}$ and the size of a $^{13}$CO core, the central density
$\rho_{c}$ is determined. The critical mass is then calculated by
integration to the density profile (equation \ref{beprofile}). Since
the upper limit and the lower limit of the integration are
constants, it is straightforward to show that the Bonnor-Ebert mass
$M_{\rm BE}\propto r$.

A $^{13}$CO core having mass greater than its Bonnor-Ebert mass
would be hydrostatically unstable. As can be seen in figure 12, no
such $^{13}$CO cores (assuming a FWHM line width of 1 km/s) are
found. This is in contrast to cores found in Orion, which are mostly
supercritical (and thus unstable)~\citep{core_mass_function}.

\begin{figure}[htb]
\begin{tabular}{cc}
\includegraphics[width=8cm]{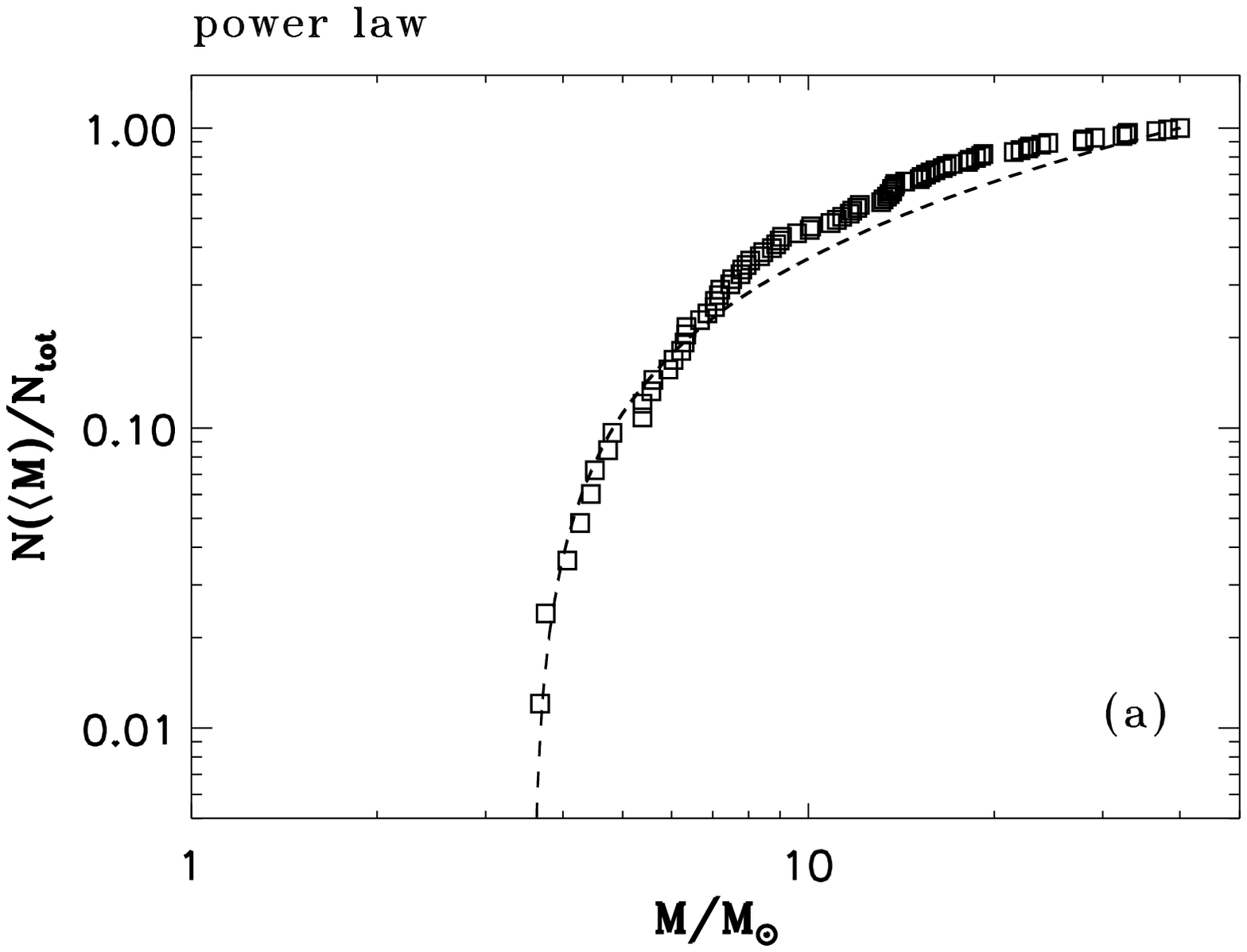} & \includegraphics[width=8cm]{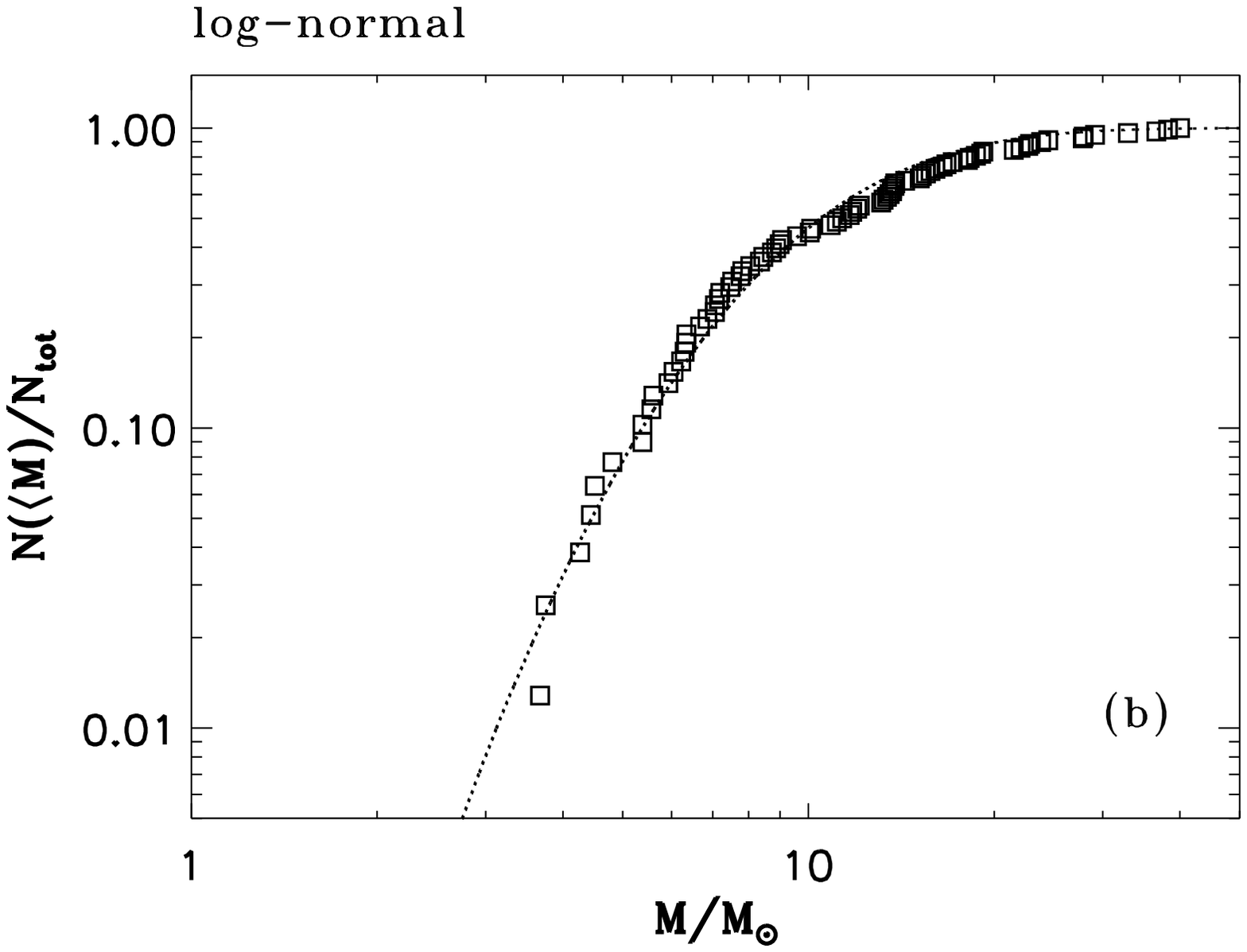}\\
\end{tabular}
\caption{ The mass distribution of those nearly gravitationally
bound $^{13}$CO cores with virial mass to mass ratio $M_{\rm
vir}/M<2$. The fitting results are (a) $dN/d\log M\propto M^{0.20}$,
$M_{\rm max}=40.04\ M_{\odot}$, $M_{\rm min}=3.57\ M_{\odot}$. (b)
log-normal distribution with $\mu=2.35$, $\sigma=0.52$. The
log-normal distribution gives a good fit over the entire mass range
that is sampled. \label{massfunction_be}}
\end{figure}

We have studied the mass distribution of nearly gravitationally
bound $^{13}$CO cores (with virial mass to mass ratio $M_{\rm
vir}/M<2$) and find that they can be well fitted with a log-normal
function with $\mu=2.35$, $\sigma=0.52$ (see
figure~\ref{massfunction_be}). The fact that a log-normal function
fits the nearly gravitationally bound cores better than an Salpeter
IMF type power law suggests that the mass conversion efficiency is
NOT constant for all $^{13}$CO cores.

We also examine the virial mass of the $^{13}$CO cores according to
Equation \ref{virial_mass}. Among the $^{13}$CO cores, there are 83
cores out of 588 that are nearly gravitationally bound (with the
virial parameter $M_{\rm vir}/M<2$, see equation~\ref{virial_mass}).
The pressure from the surrounding materials also serves as
confinement in addition to self-gravity, reducing the virial mass of
a $^{13}$CO core (e.g.~\cite{kainulainen11}). We can define an
effective mass corresponding to the pressure by calculating the
square of the effective velocity dispersion
\begin{equation}
\Delta V_{\rm eff}^2\equiv \frac{P_{\rm ext}}{\rho_{\rm core}}=\frac{kT_{\rm ext}}{m_{\rm H_2}}\frac{\rho_{\rm ext}}{\rho_{\rm core}}\lc
\end{equation}
where $T_{\rm ext}$ is the temperature of the ambient medium. An effective mass is then defined analogously to the virial mass
\begin{equation}
M_{\rm eff}=\frac{f\Delta V_{\rm eff}^2 R}{G}=\frac{fkT_{\rm ext}R}{m_{\rm H_2}G}\frac{\rho_{\rm ext}}{\rho_{\rm core}}
=4.8\left(\frac{f}{5}\right)\left(\frac{R}{0.1\rm pc}\right)\left(\frac{T_{\rm ext}}{10\rm K}\right)\left(\frac{\rho_{\rm ext}}{\rho_{\rm core}}\right)M_{\odot}\lc
\end{equation}
where $f$ is a dimensionless factor. For cores of the same mass, the pressure is a more important factor for the confinement of the less dense cores.
The external pressure may be negligible for the confinement of a dense core, for example, considering a core with $\rho_{\rm ext}/\rho_{\rm core}\sim 1/100$,
the effective mass would be about 0.05 $M_{\odot}$, which is negligible for a core having a total mass $\geq$ 1 M$_{\odot}$.
The pressure is then unimportant for the confinement
of massive cores, but is important for a core having total mass $\lesssim$ 0.1 M$_{\odot}$.

\subsection{The Motion of Cores}

\begin{figure}[htb]
\centering
\includegraphics[width=18cm]{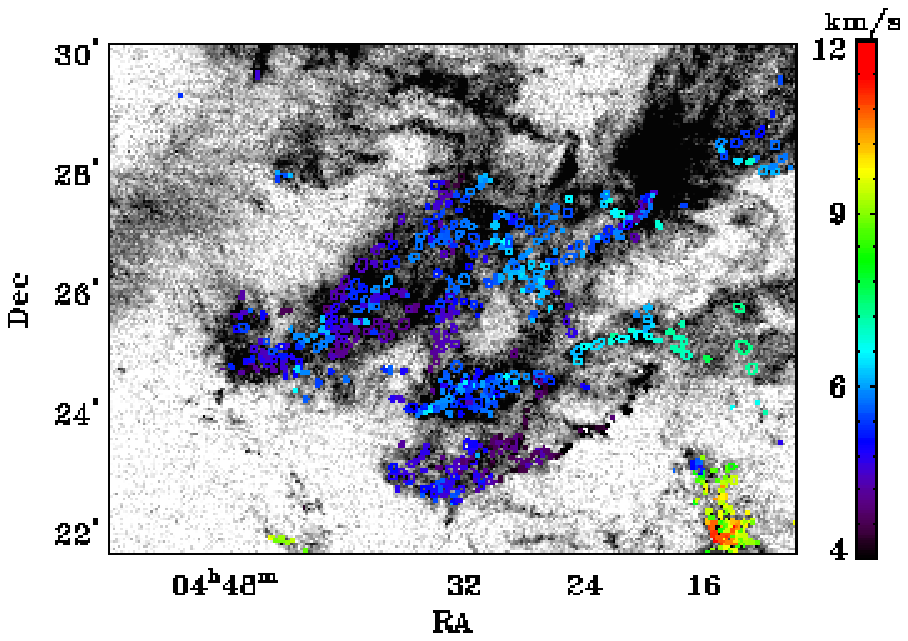}
\caption{ Overlay of the cores identified using the 3D $^{13}$CO
data cube on the $^{12}$CO total intensity map with the centroid
velocity coded as shown in the color bar on right. In most parts of
Taurus, the core centroid velocities are similar. However, at the
lower right corner, the centroid velocities of the cores differ
systematically from those in other regions, which may suggest a
different origin or location of these cores. In the study of the
core velocity dispersion, we exclude these
cores.\label{overlay_velocity}}
\end{figure}

\begin{figure}[htb]
\centering
\includegraphics[width=18cm]{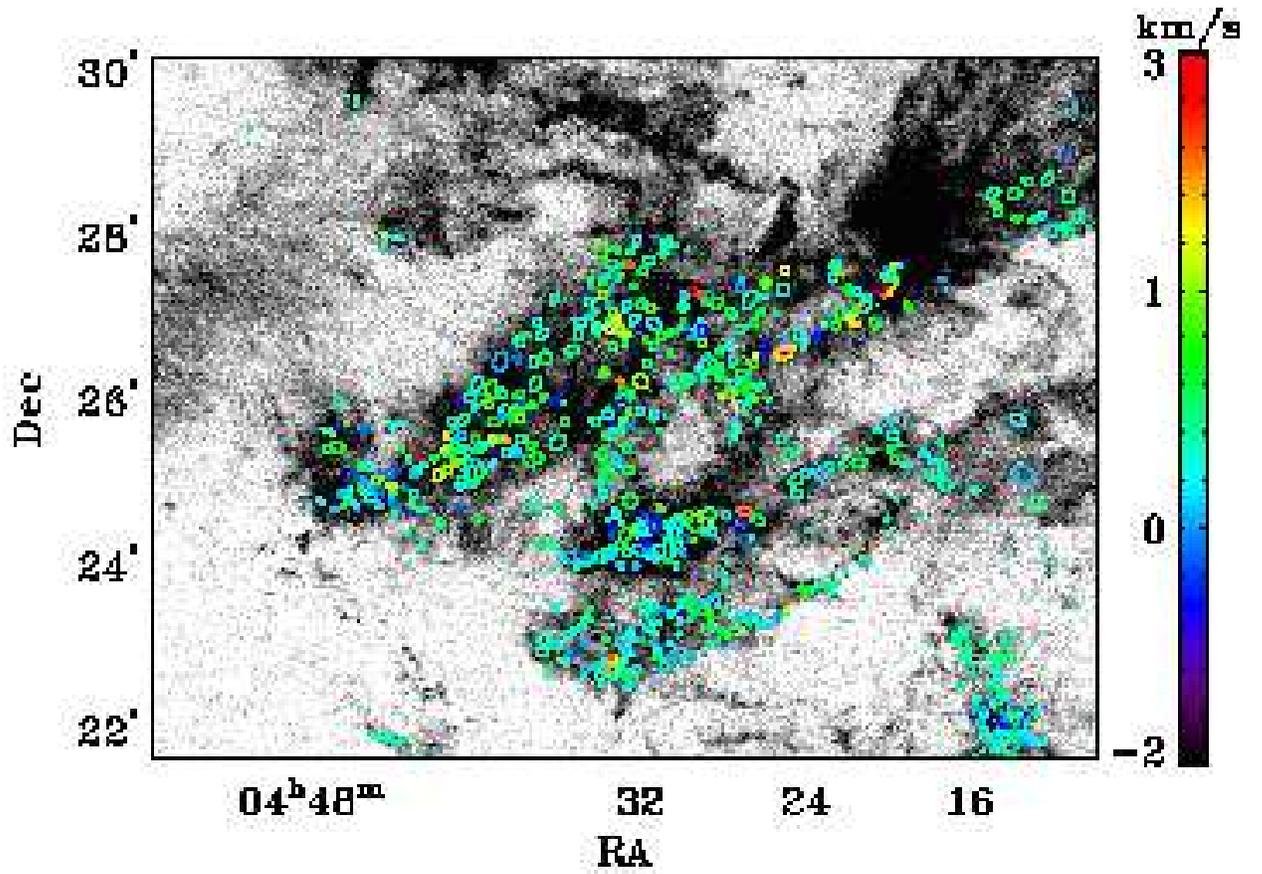}
\caption{ Overlay of the cores found in the 3D $^{13}$CO data cube
on the $^{12}$CO total intensity map. The difference between
centroid velocity of the cores and the local $^{12}$CO peak velocity
is indicated by the color (coded in the color bar on right) and is
generally small ($\leq$ 1 km/s). \label{overlay_velocity2}}
\end{figure}

\begin{figure}[htb]
\centering
\includegraphics[width=18cm]{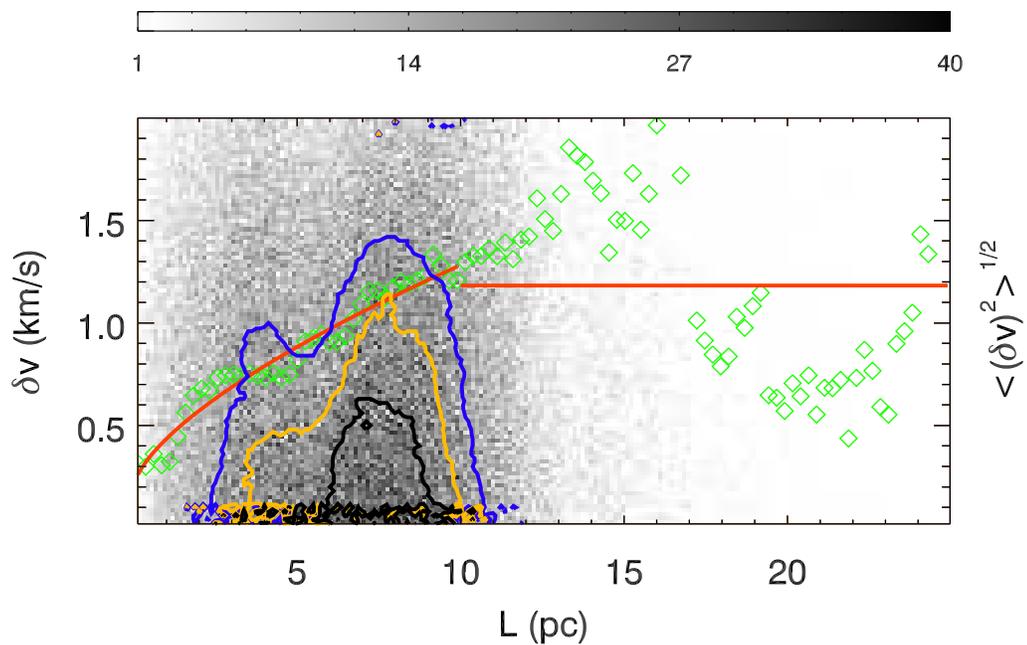}
\caption{ Plot of the core velocity difference, $\delta v$  vs.\ the
apparent separation $L$ of cores and the core velocity dispersion
(CVD$\equiv \langle\delta v^2\rangle^{1/2}$). The background is a
image of the number of data points in the $\delta v - L$ plane, with
grey scale (density bar at top of figure) showing the density of
points. Contours of the density distribution in this plot are also
shown. The green diamonds represent the variance of the velocity
difference in each separation bin. The background density
distribution shows two group of points, which are also evident by
looking at the contours. These two groups define two length scales.
One corresponds to the typical size of a "core cluster", about 4 pc,
while the other one shows the "core cluster" separation to be about
8 pc. \label{vdis_all}}
\end{figure}

Previous studies suggest that $^{13}$CO traces the bulk motion of
dense gas as well as NH$_{3}$ and N$_{2}$H$^+$
\citep{Benson1998,Kirk2010}. The cores are shown in
figure~\ref{overlay_velocity} with the fitted centroid radial
velocity coded in color. In general, the separation between the
velocity of the core and that of the ambient gas is small. As seen
in figure~\ref{overlay_velocity2}, the core centroid velocities are
also mostly similar, except for cores in the lower right corner.
This region differs systematically from the main Taurus cloud and
may not be physically connected to it. After excluding these cores,
there are 588 cores (out of 765 cores) left. We plot the velocity
difference $\delta v$ vs. the apparent separation $L$ of these cores
in figure~\ref{vdis_all}.

Between the 588 cores there are $588\times(588-1)/2$ unique pairs,
which are shown as the background density distribution in
figure~\ref{vdis_all}. The velocity difference $\delta v$ shows a
bimodal distribution. We then calculate the dispersion of the
velocity difference and plot the core velocity dispersion (CVD
$\equiv \langle\delta v^2\rangle^{1/2}$). The relation between the
CVD and the apparent separation between cores can be fitted with a
power law, yielding CVD (km/s) $=0.2L ({\rm pc})^{0.7}+0.2$ for $L$
between 0 and 10 pc, while those points having $L>10$ pc  show large
scatter, likely caused by under-sampling due to the finite size of
the cloud. The mean value of the CVD for separation $L>10$ pc is
$1.18$ km/s. The bimodal distribution in the velocity difference can
be understood as defining two spatial scales. One corresponds to the
typical size of a "core cluster", about 4 pc, while the other one
shows the typical core cluster separation, about 8 pc.

\begin{figure}[htb]
\centering
\includegraphics[width=18cm]{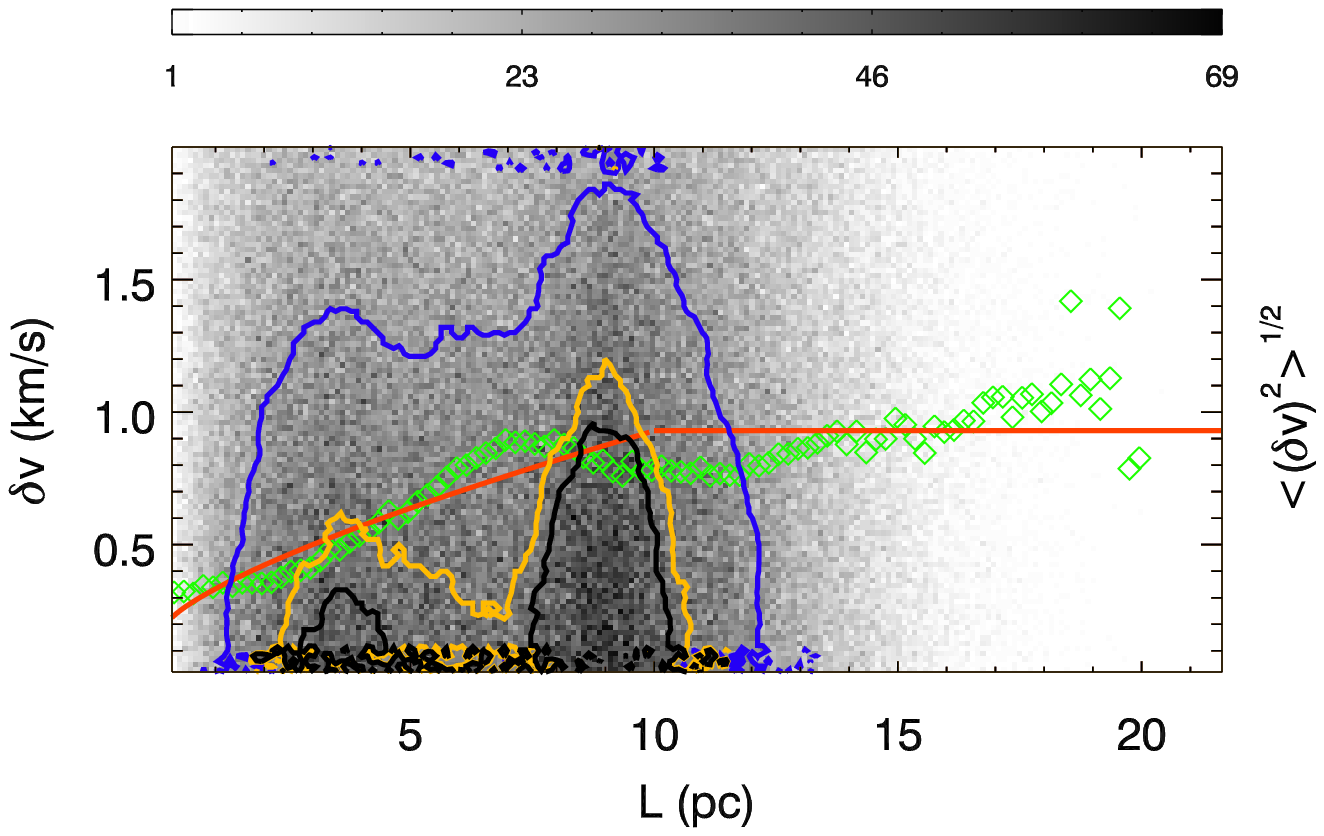}
\caption{Results from numerically generated sample taken from two
spherical clusters of cores, each having a 7 pc radius and with a
separation of 9 pc between the centers of two clusters. The same
statistical treatment as that applied to the actual data
(figure~\ref{vdis_all}) has been employed. The two length scales,
i.e. the size of a core cluster, and the separation of the clusters
are clearly seen. The points with apparent separation $ 0 \geq L
\leq 10$  pc can be fitted with a power law of CVD (km/s)$=0.13L
({\rm pc})^{0.7}+0.2$. The horizontal line shows the mean $\delta v$
value, $0.93$ km/s, of points with $L> 10$ pc.
\label{vdis_all_simu}}
\end{figure}

To evaluate the significance of the bimodal distribution, we
constructed a simple two-core-cluster model with 500 cores, with
each cluster having a radius $R=7$ pc. The clusters are separated by
9 pc. A uniform spatial distribution of cores is assumed for both
clusters. The velocity of a core is defined by a Gaussian with
variance $\sigma$, given by $(\sigma/\sigma_{\rm max}) =
(D/R)^{0.5}$, where $D$ is the distance of the core from the center
of the cluster it belongs to, and $\sigma_{\rm max}$ is the maximum
$\sigma$ at the edge of a cluster, taken here to be 1.3 km/s. In
practice, the line of sight velocity of a core is taken to be a
random number from a gaussian distribution with variance
$\sigma/\sqrt{3}$. There is an assumed systematic difference between
the line of sight velocities of the two clusters, which is 1.8 km/s
in our model. We are able to reproduce the major features of the
observed CVD plot (figure~\ref{vdis_all}) with this simple model.
Rigorously speaking, the velocity of cores in each cluster do not
follow the exact form of  Larson's law, because there is a center.
However, the effect of such a distinction is negligible. We have
tried to generate a sample of cores in which the velocity of a core
is still gaussian distributed whose variance depends on its distance
to the previously generated core, denoted as $l$, with the same
relation $\sigma \propto l^{0.5}$, and get similar results to those
shown in figure~\ref{vdis_all}. The two length scales, i.e. the size
of a core cluster, and the separation between the clusters are
clearly seen in figure~\ref{vdis_all_simu}. The relation between the
CVD and the apparent separation in the $L=0\sim 10$ pc region can be
fitted by a power law of the form CVD (km/s) $=0.13L ({\rm
pc})^{0.7}+0.2$. In $L>10$ pc region, CVD can be fitted by a linear
function of slope 0.04. The mean value of the core velocity
difference in this region is $0.93$ km/s. We conclude that the main
features of the observed CVD vs $L$ relationship and the observed
$\delta v$ are reproduced reasonably well by this simple
demonstration.

\clearpage
\clearpage

\section{Discussion}
\label{sec:discussion} The shape of the core mass function has been
for some time under scrutiny due to its possible relevance to the
origin of stellar initial mass function. We found that a log-normal
distribution best represents the $^{13}$CO CMF in Taurus, through
fitting the 3D \13co\ data cube.

\begin{figure}
\begin{tabular}{c}
\includegraphics[width=12cm]{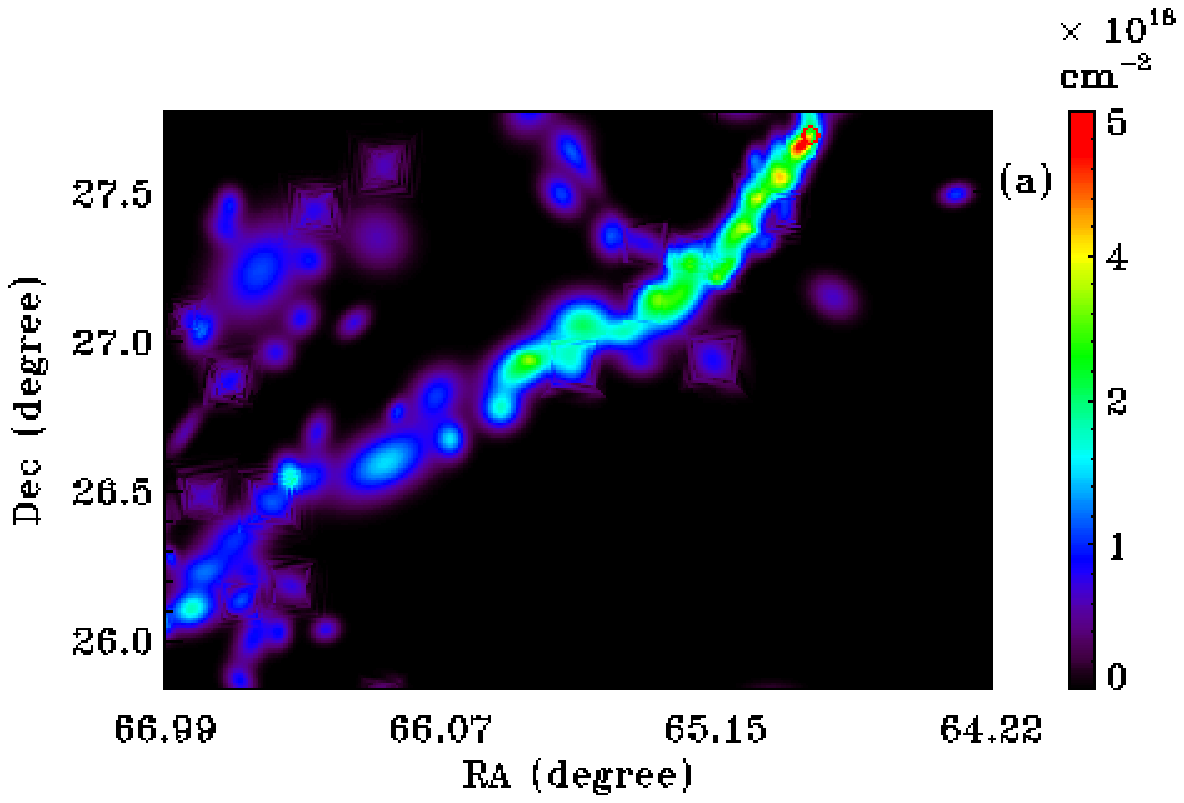}\\
\includegraphics[width=12cm]{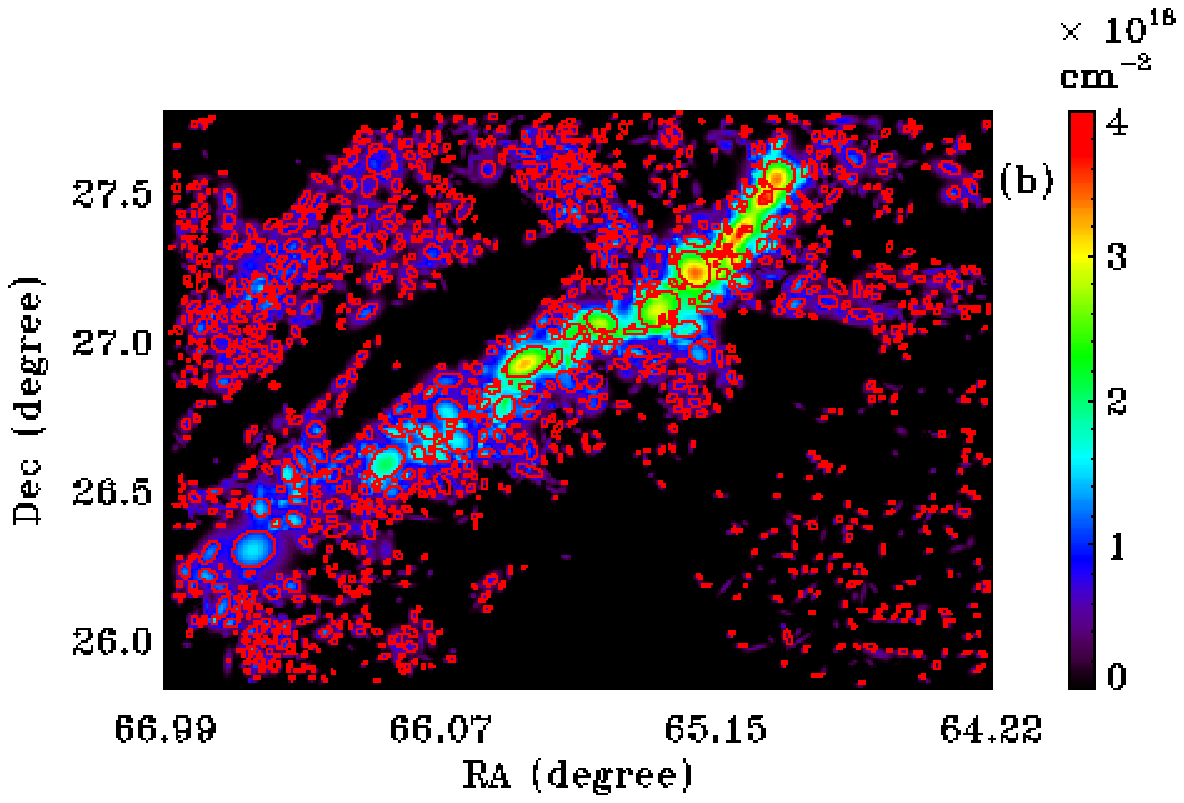}\\
\end{tabular}
\caption{Fitting to a patch within region 10. The spectrum at the
location of the red circle in the upper right of panel (a) is shown
in figure~\ref{overlap}. (a) The total intensity map of $^{13}$CO
cores found in the data cube of region 10 (3D).(b) $^{13}$CO cores
found in the total intensity of the original data of region 10 (2D).
More cores are found in the 2D fitting, especially in the relatively
diffuse regions. These cores contain large internal velocity
variations. \label{region10}}
\end{figure}

\begin{figure}
\begin{tabular}{c}
\includegraphics[width=14cm]{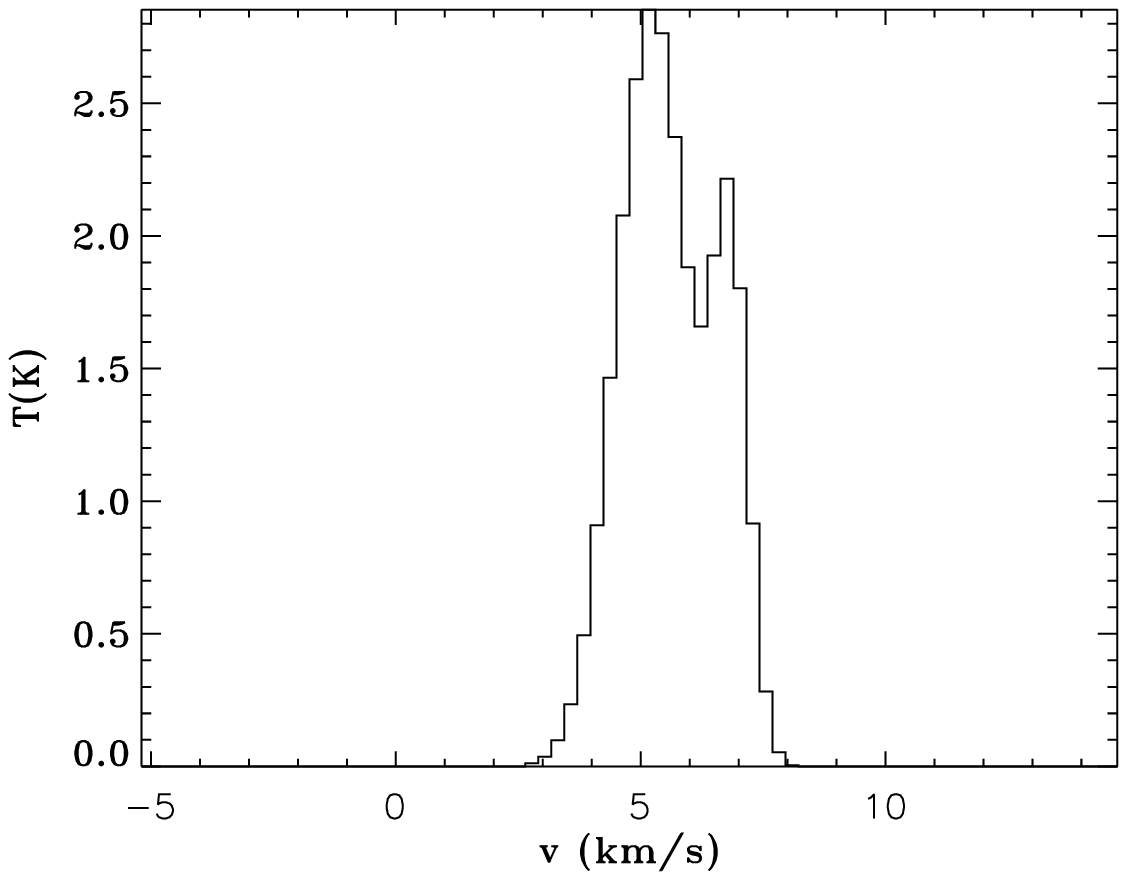}\\
\end{tabular}
\caption{An example spectrum containing contributions from two
$^{13}$CO cores, which are clearly separated in velocity. The
location where this spectrum was observed is indicated by the red
circle in panel (a) of figure~\ref{region10}. \label{overlap}}
\end{figure}

We also searched for and studied the cores identified in the 2MASS
extinction map. For extinction cores, a power law function is a
better fit than a log-normal function. It is important to note the
significant differences between 3D fitting and 2D fitting. In the
former case, velocity information is invoked so that overlapping
cores can be separated. We have compared the fitting to the 3D data
cube and that to the 2D total intensity map (figure~\ref{region10}).
There are some low-density cores that do not emerge in the fitting
to the 3D data cube. There are overlapping cores, which can be
separated in the 3D fitting by using the velocity information (an
example of the spectrum is shown in figure~\ref{overlap}). This is a
direct demonstration of the peril of solely relying on total column
density maps (such as dust continuum) to obtain core properties. In
the light of these considerations, we believe that the CMF derived
from extinction maps needs to be treated with some caution for
interpreting the origin of the IMF, even though it may have a
similar power-law form as that of the IMF.

\begin{figure}[htb]
\centering
\begin{tabular}{c}
\includegraphics[width=12cm]{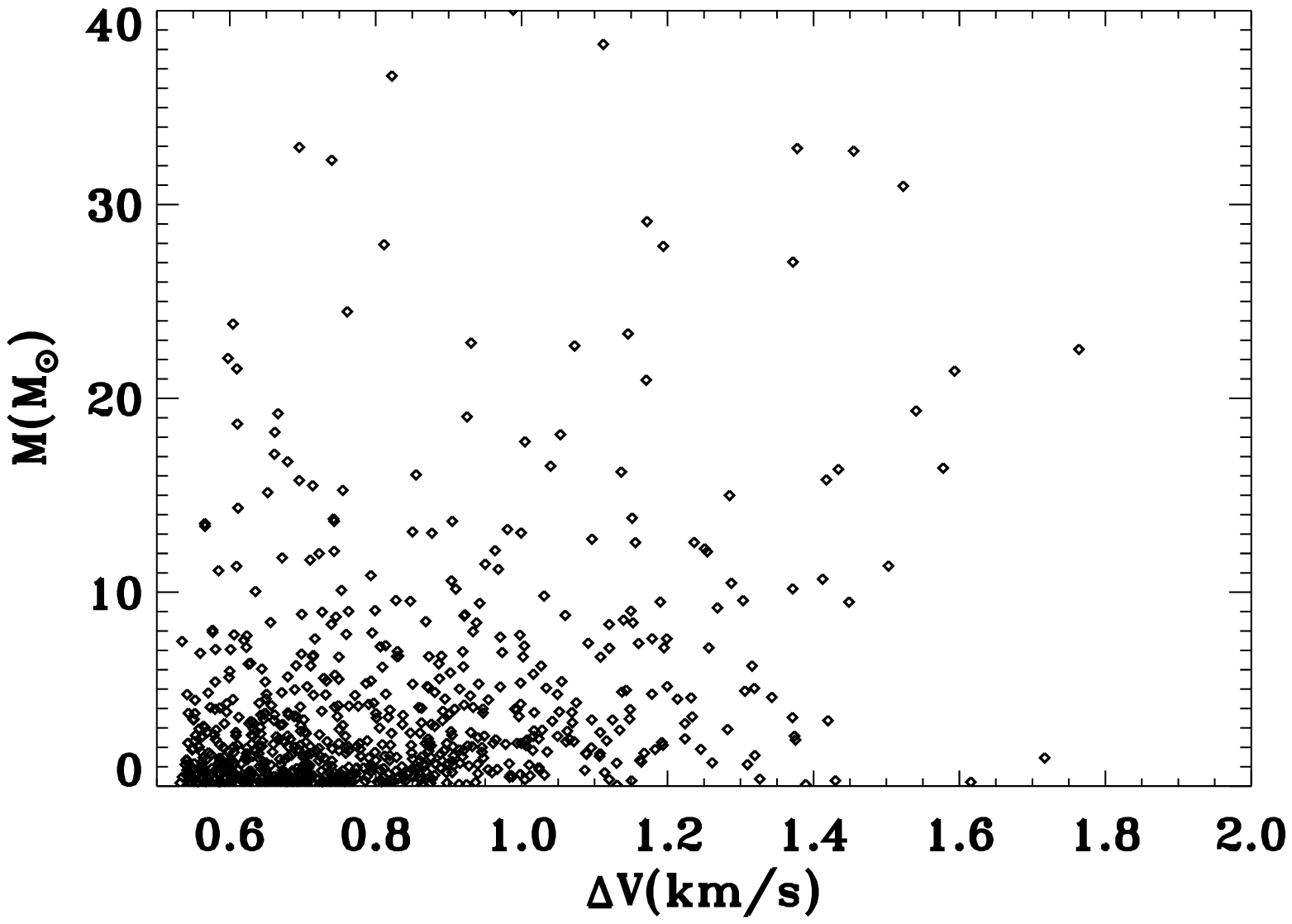}\\
\end{tabular}
\caption{ The mass-line width relation of the cores found by GAUSSCLUMPS. There is no apparent correlation between the mass and the line width.
\label{mass_velocity}}
\end{figure}

\begin{figure}[htb]
\centering
\begin{tabular}{c}
\includegraphics[width=12cm]{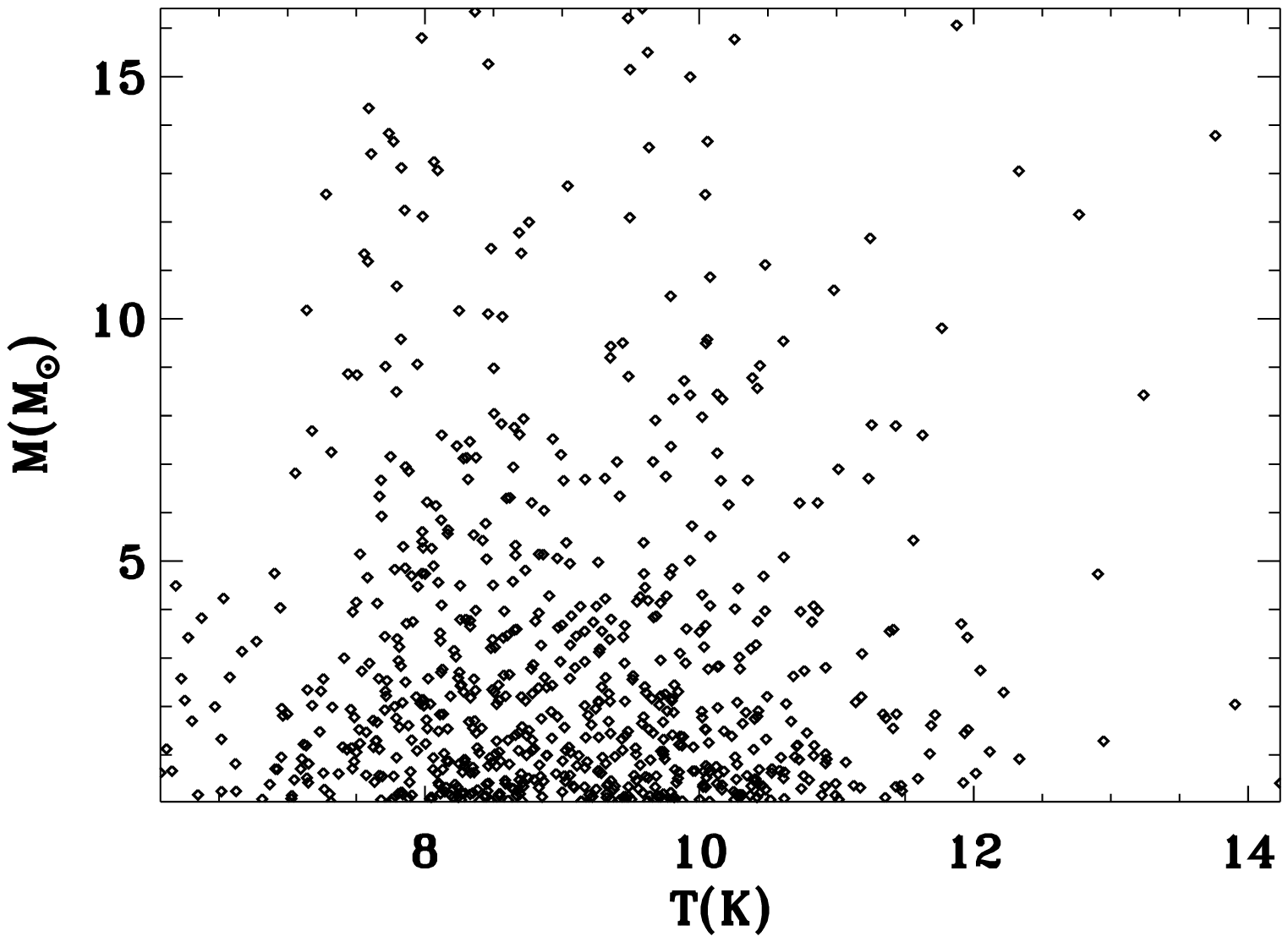}\\
\end{tabular}
\caption{The mass-temperature relation of the cores found by GAUSSCLUMPS. There is no apparent correlation between these two quantities.
\label{mass_temperature}}
\end{figure}

The large sample size and substantial spatial dynamic range of the
Taurus \13co\ core sample allow us to reconstruct the core velocity
dispersion (CVD) for the first time. The CVD exhibits a power-law
behavior as a function of the apparent separation $L$  between cores
for $L<$ 10 pc (see figure~\ref{vdis_all}). This is similar to
Larson's law for the velocity dispersion of the gas. The peak
velocities of \13co\ cores do not deviate from the centroid
velocities of ambient \co\ by more than half the \co\ line width.
The small velocity differences between dense and diffuse gas have
also been noted by ~\cite{Kirk2007} for Perseus cores. Simulation of
core formation under the influence of converging flows suggest that
massive cores exhibit relatively small line widths compared to less
massive ones~\citep{Gong2011}. We do not see this trend in our work,
i.e. there is no apparent correlation between the mass and the line
width or between the mass and the temperature (see
figures~\ref{mass_velocity} and~\ref{mass_temperature}).
\cite{Padoan2001} found smaller line width in denser gas, which is
not seen in Taurus sample either. These results suggest that dense
cores condense out of the more diffuse gas without additional energy
input from sources, such as protostars or converging flows.

In recent simulations of core formation~\citep[e.g.\ ][]{Gong2011}
or star cluster formation~\citep[e.g.][]{Offner2009}, there is
sufficient information to produce a CVD plot of the simulated core
samples. Since CVD is sensitive to the dynamic history of core
formation, we encourage the simulators to perform such analysis to
facilitate direct comparison between theoretical models and
observations.

\section{Conclusions}
\label{sec:conclusion}

We have studied the $^{13}$CO cores identified within the Taurus
molecular cloud using a 100 $\rm degree^2$ \13co\ $J=1\to 0$ map of
this region. The spatial resolution of 0.014 pc  and the velocity
resolution of 0.266 km/s facilitate a detailed study of the physical
conditions of $^{13}$CO cores in Taurus. The spatial dynamic range
(the ratio of linear map size to the Nyquist sampling interval) of
1000 of our data set allows examination of the collective motions of
$^{13}$CO cores and their relationship to their surroundings. We
have found that the velocity information helps to exclude cores
which we consider to be spurious.

Our conclusions regarding the extraction of $^{13}$CO cores and
their properties are:

1) Velocity information is essential in resolving overlapping cores,
allowing better resolution of cores and a more accurate
determination of the CMF.

2) The mass function of the 3D(x,y,v) \13co\ cores can be fitted
better with a log-normal function ($\mu = 0.95$ and $\sigma= 1.15$)
than with a power law function. For cores represented by \13co\
total intensity and extinction, a log-normal distribution is also a
better representation of the mass distribution than is a power law.
There is no simple relation between the Taurus $^{13}$CO CMF and the
stellar IMF.

3) No $^{13}$CO cores are found to have mass greater than  the
critical Bonnor-Ebert mass, in contrast to cores in Orion.

4) Only 10\% of $^{13}$CO cores are approximately  bound by their
own gravity (with the virial parameter $M_{\rm vir}/M<2$). These can
be well fitted with a log-normal mass function. External pressure
plausibly plays a significant role in confining the $^{13}$CO cores
with small density contrast to the surrounding medium.

5) In Taurus, the relation between core velocity dispersion
(CVD$\equiv \langle\delta v^2\rangle^{1/2}$) and the apparent
separation between cores $L$ can be fitted with a power law of the
form CVD (km/s) $=0.2L ({\rm pc})^{0.7}+0.2$ in the $ 0 \leq L \leq
10$ pc region, similar to the Larson's law, with a median value of
0.78 km/s.

6) The observed CVD is reproduced by using a simple two-core-cluster
model, in which there are two core clusters with radius 7 pc and a
separation 9 pc between these two clusters.

7) The low velocity dispersion among cores, the close similarity between CVD and Larson's law, and the small difference between core centroid velocities and
the ambient diffuse gas all suggest that dense cores condense out of the diffuse gas without additional energy input and are consistent with an ISM evolution
picture without significant feedback from star formation or significant impact from converging flows.

8) The CVD can be an important diagnostic of the core dynamics and
the cloud evolution. We encourage simulators to provide comparable
information based on their calculations.

\acknowledgments This work is partly supported by  China Ministry of
Science and Technology under State Key Development Program for Basic
Research (2012CB821800) and partly supported by The General Program
of National Natural Science Foundation of China (11073028). L. Qian
is partly supported by the Young Researcher Grant of National
Astronomical Observatories, Chinese Academy of Sciences and partly
supported by M. Zhu from his funding from One Hundred Person Project
of the Chinese Academy of Sciences. This work was carried out in
part at the Jet Propulsion Laboratory, operated by the California
Institute of Technology.

\clearpage

\begin{table}[htb]
\begin{center}
\caption{Parameters used in the GAUSSCLUMPS fitting of the $^{13}$CO
data cube and $^{13}$CO total intensity map.\label{gaussclumpspara}}
\begin{tabular}{|c|c|}
\hline \hline  Parameter & Value \\\hline

WWIDTH      & 2  \\\hline

WMIN        & 0.01 \\\hline

MAXSKIP & 100\\\hline

THRESH      & 5\\\hline

NPAD        & 100\\\hline

MAXBAD      & 0.05  \\\hline

VELORES     & 2  \\\hline

MODELLIM    & 0.05  \\\hline

MINPIX      & 16  \\\hline

FWHMBEAM & 2\\\hline

MAXCLUMPS   & 2147483647\\\hline

MAXNF       & 200  \\\hline \hline
\end{tabular}
\end{center}
\footnotesize \it {\rm\bf WWIDTH} is the ratio of the width of the
weighting function (which is a Gaussian function) to the width of
the initial guessed Gaussian function.

{\rm\bf WMIN} specifies the minimum weight. Pixels with weight
smaller than this value are not included in the fitting process.

{\rm\bf MAXSKIP}: If more than "{\rm\bf MAXSKIP}"  consecutive cores
cannot be fitted, the iterative fitting process is terminated.

{\rm\bf THRESH} gives the minimum peak amplitude of cores to be
fitted by the GAUSSCLUMPS algorithm. The supplied value is multipled
by the {\rm\bf RMS} noise level before being used.

{\rm\bf NPAD}: The algorithm will terminate when "{\rm\bf NPAD}"
consecutive cores have been fitted all of which have peak values
less than the threshold value specified by the "{\rm\bf THRESH}"
parameter. (From the source code cupidGaussClumps.c, one can see
that the algorithm will do the same thing when "{\rm\bf NPAD}"
consecutive cores have pixels fewer than "MINPIX".)

{\rm\bf MAXBAD}: The maximum fraction of bad pixels which may be
included in a core. Cores will be excluded if they contain more bad
pixels than this value.

{\rm\bf VELORES}: The velocity resolution of the instrument, in
channels.

{\rm\bf MODELLIM}: Model values below ModelLim times the RMS noise
are treated as zero.

{\rm\bf MINPIX}: The lowest number of pixel which a core can
contain.

{\rm\bf FWHMBEAM}: The FWHM of the instrument beam, in pixels.

{\rm\bf MAXCLUMPS}: The upper limit of the cores to be fitted. Set
to a large number so this parameter do not take effect.

{\rm\bf MAXNF}: The maximum number of evaluations of the objective
function allowed when fitting an individual core. Here it is just
set to a very large number to guarantee all the cores to be fitted.
\end{table}

\begin{table}[htb]
\begin{center}
\caption{Parameters used in the CLUMPFIND fitting of the $^{13}$CO
data cube). \label{clumpfindpara}}
\begin{tabular}{|c|c|}
\hline \hline  Parameter & Value \\
\hline DELTAT      & 5.0*RMS  \\\hline

FWHMBEAM    & 2.0     \\\hline

MAXBAD      & 0.05    \\\hline

MINPIX & 16      \\\hline

NAXIS & 3       \\\hline

TLOW        & 5*RMS   \\\hline

VELORES & 2.0 \\\hline \hline
\end{tabular}
\end{center}
\footnotesize \it
\end{table}

\begin{table}[htb]
\begin{center}
\caption{Parameters used in the GAUSSCLUMPS fitting of the
extinction map.\label{extinctionpara}}
\begin{tabular}{|c|c|}
\hline \hline

Parameter & Value \\\hline

WWIDTH      & 2\\\hline

WMIN        & 0.01  \\\hline

MAXSKIP & 50 \\\hline

THRESH & 5\\\hline

NPAD & 50  \\\hline

MAXBAD & 0.05\\\hline

VELORES     & 2 \\\hline

MODELLIM    & 0.05  \\\hline

MINPIX      & 16  \\\hline

FWHMBEAM & 2 \\\hline

MAXCLUMPS   & 2147483647\\\hline

MAXNF       & 200\\\hline \hline
\end{tabular}
\end{center}
\end{table}

\clearpage



\begin{longtable}[htb]{|c|c|c|c|c|c|c|c|c|c|c|}

\caption[Properties of the cores]{\textbf{Properties of the cores found in 3D $^{13}$CO data cube.}}\label{tab:clumps}\\

\hline \multicolumn{1}{|c|}{\textbf{ID}} &
\multicolumn{1}{|c|}{\textbf{RA($^\circ$)}} &
\multicolumn{1}{c|}{\textbf{DEC($^\circ$)}}&
\multicolumn{1}{c|}{\textbf{\begin{sideways}$\mathbf{R}_{\rm
major}(')$\end{sideways}}} &
\multicolumn{1}{c|}{\textbf{\begin{sideways}$\mathbf{R}_{\rm
minor}(')$\end{sideways}}} &
\multicolumn{1}{c|}{\textbf{$\mathbf{\theta}$($^\circ$)}}&
\multicolumn{1}{c|}{\textbf{T(K)}} &
\multicolumn{1}{c|}{\textbf{\begin{sideways}$\mathbf{M}$($\mathbf{M}_{\odot}$)\end{sideways}}}
& \multicolumn{1}{c|}{\textbf{\begin{sideways}$\mathbf{M}_{\rm
vir}$($\mathbf{M}_{\odot}$)\end{sideways}}}&
\multicolumn{1}{c|}{\textbf{\begin{sideways}FWHM(km/s)\end{sideways}}}&
\multicolumn{1}{c|}{\textbf{\begin{sideways}$\mathbf{n}_{\rm H_2, mean}$({cm}$^{-3}$)\end{sideways}}}\\
\hline

\endfirsthead

\multicolumn{11}{c}%

{{\bfseries \tablename\ \thetable{} -- continued from previous page}} \\

\hline \multicolumn{1}{|c|}{\textbf{ID}} &
\multicolumn{1}{|c|}{\textbf{RA($^\circ$)}} &
\multicolumn{1}{c|}{\textbf{DEC($^\circ$)}}&
\multicolumn{1}{c|}{\textbf{\begin{sideways}$\mathbf{R}_{\rm
major}(')$\end{sideways}}} &
\multicolumn{1}{c|}{\textbf{\begin{sideways}$\mathbf{R}_{\rm
minor}(')$\end{sideways}}} &
\multicolumn{1}{c|}{\textbf{$\mathbf{\theta}$($^\circ$)}}&
\multicolumn{1}{c|}{\textbf{T(K)}} &
\multicolumn{1}{c|}{\textbf{\begin{sideways}$\mathbf{M}$($\mathbf{M}_{\odot}$)\end{sideways}}}
& \multicolumn{1}{c|}{\textbf{\begin{sideways}$\mathbf{M}_{\rm
vir}$($\mathbf{M}_{\odot}$)\end{sideways}}} &
\multicolumn{1}{c|}{\textbf{\begin{sideways}FWHM(km/s)\end{sideways}}}&
\multicolumn{1}{c|}{\textbf{\begin{sideways}$\mathbf{n}_{\rm H_2,
mean}$({cm}$^{-3}$)\end{sideways}}}\\ \hline \hline

\endhead

\hline \multicolumn{10}{|r|}{{Continued on next page}} \\ \hline

\endfoot

\endlastfoot
\hline\hline

   1 &  4h35m 38.5s & 24d 6m 50.1s &    4.5 &    3.2 &   55.9 &   12.9 &   40 &   32 &    1.0 &       3300 \\
   2 &  4h31m 49.7s & 24d33m  2.1s &    4.9 &    3.8 &  138.9 &    9.2 &   38 &   46 &    1.1 &       3820 \\
   3 &  4h23m 33.3s & 25d 3m 22.6s &    5.8 &    3.5 &  128.3 &    9.4 &   37 &   26 &    0.8 &       2480 \\
   4 &  4h26m 50.1s & 26d13m 53.8s &    4.4 &    2.6 &   95.3 &    7.9 &   33 &   14 &    0.7 &       3130 \\
   5 &  4h31m 52.1s & 26d14m 42.7s &    5.0 &    4.9 &   18.3 &    9.2 &   33 &   80 &    1.4 &       2470 \\
   6 &  4h28m 59.7s & 24d30m 16.2s &    5.0 &    4.5 &  106.7 &    9.4 &   31 &   93 &    1.5 &       5950 \\
   7 &  4h21m  4.2s & 27d 4m 15.2s &    4.5 &    3.5 &   94.1 &    9.4 &   29 &   47 &    1.2 &       4290 \\
   8 &  4h11m 15.5s & 28d31m 31.7s &    5.3 &    3.0 &   97.4 &   10.1 &   28 &   23 &    0.8 &       4330 \\
   9 &  4h39m 42.7s & 25d45m 25.7s &    4.2 &    3.7 &  103.6 &   10.6 &   28 &   47 &    1.2 &       3680 \\
  10 &  4h39m 10.9s & 25d53m 45.0s &    3.5 &    3.3 &    3.2 &    8.3 &   27 &   55 &    1.4 &       1780 \\
\hline
\end{longtable}
{\footnotesize Properties of the 10 most massive $^{13}$CO cores.
Following the identifiers in column 1, the next two columns are the
RA and Dec of the center of the fitted $^{13}$CO cores. The
semi-major axis $R_{\rm major}$ and semi-minor axis $R_{\rm minor}$
of the fitted $^{13}$CO cores follow. The sixth column is the
position angle of $^{13}$CO cores (angle from north to the major
axis). The next column is the peak temperature of the $^{13}$CO
cores. The mass $M$ and the virial mass $M_{vir}$ follow. Column 10
gives the full width to half maximum line width of the $^{13}$CO
line. The last column shows the estimated mean $\rm H_2$ density of
the $^{13}$CO cores.}

\begin{longtable}[htb]{|c|c|c|c|c|c|c|c|c|c|}

\caption[Properties of the cores]{\textbf{Properties of the cores found in the smoothed $^{13}$CO data cube.}}\label{tab:clumps_smooth}\\

\hline \multicolumn{1}{|c|}{\textbf{ID}} &
\multicolumn{1}{|c|}{\textbf{RA($^\circ$)}} &
\multicolumn{1}{c|}{\textbf{DEC($^\circ$)}}&
\multicolumn{1}{c|}{\textbf{\begin{sideways}$\mathbf{R}_{\rm
major}(')$\end{sideways}}} &
\multicolumn{1}{c|}{\textbf{\begin{sideways}$\mathbf{R}_{\rm
minor}(')$\end{sideways}}} &
\multicolumn{1}{c|}{\textbf{$\mathbf{\theta}$($^\circ$)}}&
\multicolumn{1}{c|}{\textbf{T(K)}} &
\multicolumn{1}{c|}{\textbf{\begin{sideways}$\mathbf{M}$($\mathbf{M}_{\odot}$)\end{sideways}}}
& \multicolumn{1}{c|}{\textbf{\begin{sideways}$\mathbf{M}_{\rm
vir}$($\mathbf{M}_{\odot}$)\end{sideways}}}&
\multicolumn{1}{c|}{\textbf{\begin{sideways}FWHM(km/s)\end{sideways}}}\\
\hline

\endfirsthead

\multicolumn{10}{c}%

{{\bfseries \tablename\ \thetable{} -- continued from previous page}} \\

\hline \multicolumn{1}{|c|}{\textbf{ID}} &
\multicolumn{1}{|c|}{\textbf{RA($^\circ$)}} &
\multicolumn{1}{c|}{\textbf{DEC($^\circ$)}}&
\multicolumn{1}{c|}{\textbf{\begin{sideways}$\mathbf{R}_{\rm
major}(')$\end{sideways}}} &
\multicolumn{1}{c|}{\textbf{\begin{sideways}$\mathbf{R}_{\rm
minor}(')$\end{sideways}}} &
\multicolumn{1}{c|}{\textbf{$\mathbf{\theta}$($^\circ$)}}&
\multicolumn{1}{c|}{\textbf{T(K)}} &
\multicolumn{1}{c|}{\textbf{\begin{sideways}$\mathbf{M}$($\mathbf{M}_{\odot}$)\end{sideways}}}
& \multicolumn{1}{c|}{\textbf{\begin{sideways}$\mathbf{M}_{\rm
vir}$($\mathbf{M}_{\odot}$)\end{sideways}}}&
\multicolumn{1}{c|}{\textbf{\begin{sideways}FWHM(km/s)\end{sideways}}}\\
\hline \hline

\endhead

\hline \multicolumn{10}{|r|}{{Continued on next page}} \\ \hline

\endfoot

\endlastfoot
\hline\hline

   1 &  4h19m 31.2s & 27d 8m 60.0s &    6.8 &    4.4 &  153.9 &    8.0 &  104 &  154 &    4.3 \\
   2 &  4h21m 12.0s & 27d 2m 24.0s &   10.2 &    7.2 &  124.5 &    9.4 &   77 &  356 &    5.2 \\
   3 &  4h32m 43.2s & 24d21m 36.0s &    7.3 &    5.0 &  146.9 &   11.1 &   77 &   89 &    3.1 \\
   4 &  4h30m 52.8s & 26d53m 24.0s &    6.9 &    6.2 &   88.8 &    8.4 &   72 &   61 &    2.4 \\
   5 &  4h11m 14.4s & 28d33m 36.0s &    9.2 &    5.9 &  117.7 &   11.6 &   63 &   56 &    2.2 \\
   6 &  4h23m 38.4s & 25d 0m  0.0s &    7.9 &    5.6 &  133.7 &   10.3 &   47 &   48 &    2.2 \\
   7 &  4h31m 52.8s & 26d17m 60.0s &    6.0 &    5.2 &    6.5 &   10.0 &   35 &   94 &    3.3 \\
   8 &  4h32m 43.2s & 26d 2m 60.0s &    6.0 &    5.1 &  132.8 &    9.1 &   34 &  236 &    5.3 \\
   9 &  4h31m 43.2s & 24d31m 48.0s &    5.7 &    5.6 &  152.5 &   11.4 &   34 &  265 &    5.5 \\
  10 &  4h35m 45.6s & 22d56m 24.0s &    5.3 &    4.0 &   71.3 &    6.8 &   31 &   48 &    2.6 \\
\hline
\end{longtable}
{\footnotesize The columns are arranged as those of
table~\ref{tab:clumps}. }

\begin{center}

\clearpage
\begin{longtable}[htb]{|c|c|c|c|c|c|c|}

\caption[Properties of the cores]{\textbf{Properties of the cores found in extinction map.}}\label{tab:clumpsEx}\\

\hline \multicolumn{1}{|c|}{\textbf{ID}} &
\multicolumn{1}{|c|}{\textbf{RA($^\circ$)}} &
\multicolumn{1}{c|}{\textbf{DEC($^\circ$)}}&
\multicolumn{1}{c|}{\textbf{\begin{sideways}$\mathbf{L}_{\rm
major}(')$\end{sideways}}} &
\multicolumn{1}{c|}{\textbf{\begin{sideways}$\mathbf{L}_{\rm
minor}(')$\end{sideways}}} &
\multicolumn{1}{c|}{\textbf{$\mathbf{\theta}$($^\circ$)}} &
\multicolumn{1}{c|}{\textbf{\begin{sideways}$\mathbf{M}$($\mathbf{M}_{\odot}$)\end{sideways}}}\\
\hline

\endfirsthead

\multicolumn{7}{c}%

{{\bfseries \tablename\ \thetable{} -- continued from previous page}} \\

\hline \multicolumn{1}{|c|}{\textbf{ID}} &
\multicolumn{1}{|c|}{\textbf{RA($^\circ$)}} &
\multicolumn{1}{c|}{\textbf{DEC($^\circ$)}}&
\multicolumn{1}{c|}{\textbf{\begin{sideways}$\mathbf{R}_{\rm
major}(')$\end{sideways}}} &
\multicolumn{1}{c|}{\textbf{\begin{sideways}$\mathbf{R}_{\rm
minor}(')$\end{sideways}}} &
\multicolumn{1}{c|}{\textbf{$\mathbf{\theta}$($^\circ$)}} &
\multicolumn{1}{c|}{\textbf{\begin{sideways}$\mathbf{M}$($\mathbf{M}_{\odot}$)\end{sideways}}}\\
\hline

\endhead

\hline \multicolumn{7}{|r|}{{Continued on next page}} \\ \hline

\endfoot

\endlastfoot
\hline\hline

   1 &  4h40m28.8s & 25d30m 0.0s &  10.7 &   4.9 &   94.8 &  89 \\
   2 &  4h18m24.0s & 28d26m24.0s &   4.8 &   4.4 &    6.3 &  82 \\
   3 &  4h39m14.4s & 25d52m48.0s &   5.7 &   3.6 &  149.7 &  63 \\
   4 &  4h33m52.8s & 29d34m48.0s &  17.8 &   8.4 &   84.2 &  62 \\
   5 &  4h13m50.4s & 28d13m12.0s &   5.5 &   2.6 &  134.3 &  58 \\
   6 &  4h38m 2.4s & 26d14m24.0s &   8.9 &   3.5 &  124.0 &  55 \\
   7 &  4h39m40.8s & 26d10m12.0s &   6.0 &   3.9 &  143.0 &  50 \\
   8 &  4h29m19.2s & 24d33m36.0s &   5.4 &   3.1 &  123.5 &  49 \\
   9 &  4h40m55.2s & 25d54m36.0s &   7.0 &   2.3 &  159.2 &  43 \\
  10 &  4h16m60.0s & 28d40m12.0s &   7.3 &   4.8 &  113.1 &  41 \\
\hline

\end{longtable}
{\footnotesize Properties of the 10 most massive cores found in
2MASS extinction maps. Following the core identifiers, the next two
columns are the RA and Dec of the center of the fitted cores, the
the semi-major $R_{\rm major}$ and semi-minor $R_{\rm minor}$ of the
fitted cores. The sixth column is the position angle of cores (angle
from north to the major axis).  The last column is the core mass
$M$.}

\end{center}

\end{document}